\documentclass[pre,floatfix,twocolumn, 10pt]{revtex4-2} 
\usepackage{graphicx}
\usepackage{color}
\usepackage{bm}
\usepackage[colorlinks=true, linkcolor=blue, citecolor=blue]{hyperref}
\usepackage{amssymb,amsfonts,amsmath}
\usepackage{cleveref}

\newcommand{\br}{{\bm r}}
\newcommand{\sigmaY}{{\sigma_{\tt Y}}}
\DeclareMathOperator{\sgn}{sgn}

\newcommand{\resub}[1]{{#1}}

\begin{document}

\title{Thermal activation drives a finite-size crossover from scale-free to runaway avalanches in amorphous solids}

\title{Thermal activation drives a finite-size crossover from scale-free to runaway avalanches in amorphous solids}

\author{Gieberth Rodriguez-Lopez} 
\affiliation{Instituto de Física del Sur (IFISUR), Universidad Nacional del Sur (UNS), CONICET, 
12 de Octubre 1865, Bahía Blanca, Argentina.}
\affiliation{Instituto de Nanociencia y Nanotecnolog\'{\i}a, CNEA--CONICET, 
Centro At\'omico Bariloche, R8402AGP S. C. de Bariloche, 
R\'{\i}o Negro, Argentina.}
\author{Ezequiel E. Ferrero}
\affiliation{Instituto de Nanociencia y Nanotecnolog\'{\i}a, CNEA--CONICET, 
Centro At\'omico Bariloche, R8402AGP S. C. de Bariloche, 
R\'{\i}o Negro, Argentina.}

\date{\today}

\begin{abstract}
We investigate thermal avalanche dynamics in amorphous solids using 
elastoplastic models with local activation rules and no external driving. 
Dynamical heterogeneities, quantified through persistence measurements 
and the associated four-point susceptibility $\chi_4$, reveal the emergence 
of correlated spatiotemporal rearrangements as temperature is varied. 
As temperature increases, avalanche statistics evolve from 
scale-free behavior with exponential cutoffs to regimes dominated by 
system-spanning runaway events. 
We identify a system-size-dependent critical temperature $T_c(L)$ 
that separates intermittent avalanche dynamics from thermally assisted flow, 
where self-sustained avalanches transiently fluidize the system. 
We show that $T_c(L)$ decreases algebraically with increasing system size, 
suggesting that in the thermodynamic limit arbitrarily small but finite 
temperatures may destabilize the intermittent regime. 
The relation between avalanche size and duration resembles that in sheared systems, 
whereas the statistics of minimal distances to yielding reveal a temperature-driven 
reorganization of marginal stability absent in strictly driven overdamped dynamics.
Our results demonstrate that thermal activation alone can generate 
a finite-size-controlled instability scale in disordered elastic media.
\end{abstract}

\maketitle

\section{Introduction} 

Driven disordered systems -such as elastic interfaces moving in random media
and amorphous solids under deformation- display a punctuated dynamics in the 
vicinity of their critical thresholds.
Close to the pinning/depinning transition of an elastic interface driven
on a disordered landscape, the dynamics is tortuous: 
pinned regions of the interface are larger and larger as one approaches 
the critical point from above and so are the portions of the manifold 
that need to cooperate in order to move it forward.
The dynamics is commonly described as proceeding through ‘avalanches’
(spatio-temporal correlated reorganizations of the system).
Avalanches, quantified by the areas of these correlated regions or 
the length of the interface involved in the reorganization, 
tend to display very broad power-law statistical distributions of sizes 
and durations, with (presumably) universal power-law exponents 
characterizing them.
Furthermore, the exponent characterizing the avalanche size distribution
fulfills scaling relations with other critical exponents of the problem
such as the diverging length and roughness
exponents~\cite{ChauvePRB2000,KoltonPRL2006,RossoPRB2009,FerreroCRP2013,FerreroARCMP2021}.
Similarly, when an amorphous material is sheared at slower and slower strain rates,
approaching the quasistatic limit and so its dynamic yield stress, 
a similar avalanche statistics can be derived from the plastic activity 
taking  place in the system.
Essentially, due to the applied deformation the stress loads in the 
system, eventually a region of the material cannot bear more stress 
and a shear transformation occurs locally, stress is relaxed in the region 
and redistributed elsewhere, such event can cause other regions to 
locally yield, and a correlated cascade of such plastic events 
constitutes an `avalanche'. 
Experimentally, the avalanche magnitude can be quantified by the global 
stress drop that it produces, by simply looking at the stress-strain signal.
The avalanche size (and duration) statistics close to the yielding 
transition are also power-law distributed and hold finite 
system scaling laws, and exponent scaling 
relations~\cite{LinPNAS2014, liu2016, NicolasRMP2018, FerreroSM2019}, 
just as in depinning~\cite{FerreroPRL2019}.

In athermal systems, power-law distributed avalanches are only observed 
close to the critical point, both in elastic interfaces and in amorphous 
solids, and transition to exponential distributions as the driving increases
away from the critical thresholds~\cite{liu2016}.
Below threshold, there is no avalanche activity in a steady state (only transient).
Nevertheless, when a finite temperature is considered,
movement (or plastic activity) can take place below threshold and a steady 
state established.
In the subcritical region of the depinning phase diagram, in fact 
at vanishingly small forces, one finds the well known
`activated creep' regime, at small but finite $T$ and $f$,
where one can study avalanche statistics~\cite{KoltonPRL2006, KoltonPRB2009, FerreroPRL2014, FerreroARCMP2021}.
\textit{But what happens strictly at zero force and finite temperature?}
A key question is whether purely thermal activation can generate an 
intrinsic instability scale organizing avalanche dynamics in the absence 
of external driving.


In the last few years, there has been a growing interest in the 
glass community in the use of coarse-grained elastoplastic models 
(EPMs)~\cite{NicolasRMP2018}, originally proposed to study the 
rheology of yield stress systems, to analyze the problem 
of facilitation and thermal 
avalanches~\cite{OzawaPRL2023, TaheiPRX2023, TakahaPRE2025, deGeusPRE2025, KorchinskiPRX2025}.
Ozawa and Biroli~\cite{OzawaPRL2023} showed that EPMs were useful to 
reproduce the facilitation mechanism of glass-forming liquids dynamics.
They devised a classical two-dimensional EPM  where local relaxation events are 
induced by thermal fluctuations, through activation over local energy barriers
$\Delta E \propto (\sigma_{Y}-|\sigma_{i}|)^a$, 
where $\sigma_{Y}$ are the local stress thresholds for yielding, and $a$ is 
some exponent, usually $a=3/2$~\cite{OzawaPRL2023, RodriguezLopezPRM2023}.
\resub{In this work we use $\sigma_{Y}=1$ for all blocks
(notice that a site becomes unstable either when $\sigma_i>\sigma_Y$
or $\sigma_i<-\sigma_Y$.}
Since an external shear driving is completely absent, the model also includes
a randomization for the orientation angle $\varphi$ of the Eshelby propagator 
at each plastic event.
Each plastic event produces a long-range, elasticity-mediated, distortion
in the rest of the system and can therefore `induce' or `facilitate' new 
events by bringing the local stresses closer or above their local thresholds. 
The authors showed that the model displayed a relaxation time diverging
with $1/T$ and a two-point correlation function $\xi_4$ with a peak shifting
to longer times and larger amplitudes as temperature is decreased. 
This, together with snapshots of local persistence, demonstrated that these simple
EPMs were able to reproduce salient features of the physics of the glass transition, as the existence of dynamical heterogeneities and the emergence of dynamical correlations.

Following~\cite{OzawaPRL2023}, Tahaei \textit{et al.}~\cite{TaheiPRX2023} explored
the statistical properties of the EP model including avalanche statistics and 
re-discussed the facilitation mechanisms leading to dynamical heterogeneities.
They established an extremal dynamics ($T=0^{+}$) to study `thermal avalanches' and
interpret results both on avalanche statistics and dynamical heterogeneities
in terms of criticality associated to a zero-temperature fixed point. 

Korchinski~\textit{et al.}~\cite{KorchinskiPRX2025} presented a microscopic 
description of aging and logarithmic creep driven by thermal avalanches in 
elastoplastic models and Mylar sheets, pretty much in line with 
exhaustion theory~\cite{CottrellJMPS1952}. 
The gradual evolution of the distribution of local energy barriers, 
progressively develops an increasing energy gap over time,
explaining the aging and logarithmic creep.
For the elastoplastic model implementation of thermal avalanches
they introduce a finite activation probability $\lambda(x)=\exp[-E/T]/\tau_0$
for stable blocks ($x>0$), with $E=\epsilon_0 x^\alpha$.
Yet, they stay in a limit comparable with extremal dynamics by using a 
single very low temperature $T=10^{-3}$ and avoiding finite-size-induced
thermal flows ($T<T^*(L)$\footnote{Here $T^*(L)$ denotes the characteristic temperature 
(for a given system size $L$) at which we observe a crossover to a different dynamical regime.}
) by staying in moderate small system sizes.
Interestingly, for the distribution of thermal avalanches in this limit
they find $P(S)\sim S^{-\tau}$ with $\tau\simeq 1.5$.

More recently, De Geus~\textit{et al}\cite{deGeusPRE2025} have extended the 
discussion on dynamical heterogeneities to the case of elastic manifolds
driven in disordered systems, this is, to the framework of the depinning 
transition~\cite{ChauvePRB2000, KoltonPRL2006, KoltonPRB2009, RossoPRB2009, FerreroCRP2013, FerreroARCMP2021}.
The length-scale associated with the cutoff of the thermal avalanches there 
is not only related with a zero-temperature critical point but also depends
on the driving force $\ell_c(T,f)\propto T^{-\sigma}f^{-\lambda}$.
The authors provide theoretical and numerical arguments to indicate that
$\sigma=\nu$. 
This is, the way in which $\ell_c$ diverges when $T\to 0$ is essentially
the same way in which $\ell_{\tt dep} \sim |f-f_c|^{-\nu}$ diverges when 
we approach $f_c$.
$\ell_{\tt dep}$ measures the length of depinning avalanches at zero temperature
while, in the former case, $\ell_c(T,f)$ quantifies the scale of thermal avalanches.

Beyond elastic manifolds and elastoplastic models, 
Takaha \textit{et al.}~\cite{TakahaPRE2025} investigated plastic 
rearrangement events in quiescent glasses at finite temperatures 
after a sudden quench, by using molecular dynamics (MD) simulations. 
As in the driven case, local plastic rearrangements can induce others
as they produce displacement fields analogous to the elastic Eshelby field. 
The study shows, in agreement with the observations in EPMs that the size distribution
of these `thermal avalanches' follows a power-law, indicating critical behavior. 
Nevertheless, the exponents found for the avalanche size statistics ($P(S)\propto S^{-\tau}$, $\tau \approx  1$) differ from those reported by~\cite{TaheiPRX2023}.

In the present study, we set dynamical rules that incorporate
temperature through Arrhenius activation. 
Therefore, we go beyond the extremal dynamics case, 
studying thermal avalanches at `truly' finite temperatures.
As in previous works, we employ an elastoplastic model 
with a thermal activation protocol but no external driving.
We find that for each system size $L$ there is a `critical' 
temperature $T_c(L)$ above which the avalanches can self-sustain,
for transient but very long periods of time (as long as desired
if one keeps increasing the temperature), 
representing a sort of transition to a `fluidized' or `melted' state.
Moreover, $T_c(L)$ is seen to decrease with the system size,
implying that in the thermodynamic limit the solid-like phase, 
where avalanches of plastic events take place but eventually stop,
completely vanishes at any finite temperature. 

\section{Elastoplastic model and dynamical protocol}
\label{sec:epm} 

We implement a coarse-grained elastoplastic model \resub{(EPM)}
of disordered solids~\cite{RevModPhys_EPM}. 
A two-dimensional amorphous material is described by a scalar stress field 
$\sigma(\br,t)$ at position $\br\equiv(x,y)$ on a square lattice of size $N=L^{2}$ 
at time $t$. 
Let us represent the position of each cell or block of the lattice by a single 
index ``$i$'' or ``$j$''.
If at a given moment the stress $\sigma_j$ on block $j$ reaches a local 
threshold $\sigma_{Y}$, 
the block `yields': relaxing locally and 
redistributing stress in the rest of the system through a long-range
interaction kernel $G_{ij}$.
Moreover, since there is no external strain breaking the symmetry, one needs 
to consider the possibility of yielding in the equally equivalent `negative' 
loading direction, this is, to consider also the negative 
threshold $-\sigma_{Y}$ for deterministic local yielding.
Now, when a finite temperature is taken into account, a block can yield 
stochastically even without reaching its thresholds due to activation.

The model more concretely defined by a set of rules, as follows.
A block yields deterministically if its stress overcomes in absolute 
value the threshold ($\left| \sigma_{i} \right| \geq \sigmaY$), 
and also every block with $\left| \sigma_{i} \right| < \sigmaY$ 
has a finite probability per unit time of yielding due to thermal 
activation (two different kinds of activation rules defined 
in Secs.~\ref{sec:thermaldynamics} and \ref{sec:extremaldynamics} below).
When a block yields we choose the relaxation to be instantaneous, 
$\sigma_i \to \sigma_i - \delta\sigma_i$, following~\cite{TaheiPRX2023}, 
where $\delta\sigma_i$ is the local stress-drop computed as
$\delta\sigma_i = \left(z+ \left| \sigma_{i}\right| -\sigmaY \right)\sgn(\sigma_{i})$, 
with $\sgn$ is the sign function.
The stress drop tries to bring $\sigma_i$ to zero, subtracting or adding 
its previous value, but adding also some stochasticity by drawing $z$ as 
a random number from an exponential distribution of the 
form $p(z)=\frac{1}{z_{0}}e^{-z/z_{0}}$, with $z_{0}=1.0$.
Notice that, on average, $\left<z-\sigma_{Y}\right>=0$.
The local yielding of block $i$  also affects the stress value in all other 
blocks $j \neq i$, through
\begin{equation}
    \sigma_j \to \sigma_j + G_{ij} {\delta\sigma}_i.
    \label{eq:stress_evolution}
\end{equation}

Let us then introduce the propagator $G_{ij}$, which would be 
a `random-orientation' Eshelby kernel in this case, due to the 
absence of a preferential direction of deformation.
We start by recalling the Eshelby propagator derived for the effect of 
a localized plastic event in a globally elastic system~\cite{PicardAjLeBo-EPJE2004},
 \begin{equation}
     G(r, \theta) \equiv  \frac{\cos(4\theta)}{\pi r^{2}},
 \end{equation}
where $r=|\textbf{r}-\textbf{r}'|$ is the distance from the observation point 
$\textbf{r}'$ to the plastic event $\textbf{r}$ and $\theta$ the angle between the 
direction of deformation and the vector $\textbf{r}-\textbf{r}'$, 
$\theta = \arccos \left(\left(\textbf{r}-\textbf{r}'\right).\textbf{r}\right)$.
It is convenient to assume periodic boundary conditions and 
write the propagator in Fourier space, usually presented as:
\begin{equation}
    \mathcal{G}(\textbf{q}) = -\frac{(q_{x}^{2}-q_{y}^{2})^{2}}{\left(q_{x}^{2}+q_{y}^{2}\right)^{2}} ,
    \label{Eq.clasic_propagator}
\end{equation}
\noindent for $q \neq 0$ and $\mathcal{G}(\textbf{q} = \textbf{0})= -1$, 
with $q_{x}= \frac{2\pi n_{x}}{L}$, $q_{y}= \frac{2\pi n_{y}}{L}$ and 
$n_{x}, n_{y} = \left\{ -\frac{L}{2},.., \frac{L}{2} \right\}$.
By using $q_{x,y}^2 = 2-2\cos(\pi n_{x,y}/L)$, an alternative representation 
of this propagator is
\begin{equation}
    \mathcal{G}(q)= -\frac{4\left(1-\cos(q_{x})\right)\left(1-\cos(q_{y})\right)}{\left(2-\cos(q_x)-\cos(q_y)\right)^2} .
    \label{Eq.Rotate_propagator}
\end{equation}
If we now consider a rotation of an angle $\varphi \in [0, \pi/2]$,
we can write
\begin{equation}
    \mathcal{G}(q, \varphi)= -\frac{4\left(1-\cos(q_{x,\varphi})\right)\left(1-\cos(q_{y,\varphi})\right)}{\left(2-\cos(q_{x,\varphi})-\cos(q_{y, \varphi})\right)^2} 
    \label{Eq.Random_Rotate_propagator}
\end{equation}
with, 
\begin{eqnarray}
    q_{x,\varphi} =  \cos(\varphi)q_x  + \sin(\varphi)q_y, \\
    q_{y,\varphi} = -\sin(\varphi)q_x +  \cos(\varphi)q_y. 
\end{eqnarray}
We use for $G_{ij}$ in \eqref{eq:stress_evolution} essentially the
propagator \eqref{Eq.Random_Rotate_propagator} with a random rotation 
$\varphi$ uniformly distributed in $[0, \pi/2]$.
In this work, we take $\Delta t$ as the fundamental unit of time, 
with the duration a plastic event corresponding to such a time step.
At each step $t$ we draw a single random rotation angle $\varphi(t)$, 
and use the corresponding kernel $G(q,\varphi(t))$ to compute the stress redistribution from all plastic events occurring during that step. Within a time step, the propagator orientation is constant systemwide and is redrawn independently between steps.

In the following, we describe the dynamics.
We use two different protocols: a classical `thermal dynamics'
and a so-called `extremal dynamics'.

\subsection{Thermal dynamics}
\label{sec:thermaldynamics}

As mentioned, a block yields deterministically if its stress 
overcomes in absolute value the threshold 
($\left| \sigma_{i} \right| \geq \sigmaY$), but, 
due to thermal activation,
every block with $\left| \sigma_{i} \right| < \sigmaY$ 
has also a finite probability of yielding.
Considering a finite temperature $T>0$, we use an Arrhenius-type 
law to estimate such activation probability per unit time, i.e, 
\resub{
\begin{equation}\label{eq:prob}
p_{\text{act}}(T)=\frac{1}{\tau_0}e^{-B\left( \sigmaY-|\sigma_{i}|\right)^{\alpha}/k_{B}T},
\end{equation}
}
\noindent where we choose \resub{$k_{B}=\tau_0=B=1$} and $\sigmaY=1$. 
$\alpha$ is a parameter mimicking the shape of the energetic 
barriers over which activation takes 
place~\cite{FerreroSM2019, FerreroPRM2021, PopovicPRE2021}, 
and in this work we use $\alpha=3/2$.
\resub{The prefactor $B=[{\tt energy}]/[{\tt stress}]^\alpha$ in the exponential is 
introduced in order to keep the correct units~\cite{RodriguezLopezPRM2023}.}

Now, at any finite temperature $T$, plastic activity can  
occur `spontaneously' at any time and we shall observe it if we wait 
long enough iterating our dynamics.
Yet, since we work at low temperatures and we want to be computationally
more efficient, we take a small protocol license:
Starting from a situation where all blocks are mechanically stable, this is,
$\left| \sigma_{i} \right| < \sigmaY ~,~ \forall i$, we first detect which is 
the closest site to instability
(in other words, we search for $x_{\tt min} = \min_i (\sigmaY- |\sigma_{i}|)$
and identify its position $i'$). 
Then we simply `activate' it, producing the immediate fluidization of block $i'$.
The consequent stress redistribution can cause other blocks to yield, thereby 
triggering a cascade effect, an `avalanche'.

After that `first' activation, on each time step of the dynamics the stresses evolve 
according to Eq.~\ref{eq:stress_evolution} (concurrently, in parallel) and we check 
the resulting stresses of all blocks and decide whether some of them should yield
in the new time step.
Be them either deterministic yieldings (blocks getting over-stressed due to the 
stress redistribution) or thermal activations with the probability (\ref{eq:prob}) 
at temperature $T$, every new fluidized block is considered to belong to the 
same avalanche under way. 
The cascade of plastic events goes on until, at a given time step, 
not a single block yields.
In our definition, this marks the end of an avalanche.
We then trigger a new one by checking again for $x_{\tt min}$ and 
`manually' activating the block where it occurs. 
When monitoring time, we add to the time variable an amount inversely 
proportional to the \textit{a priori} probability for such an event to 
occur spontaneously~\footnote{As per Eq.~\ref{eq:prob},
$\tau_w=\tau \exp[x_{\tt min}^{3/2}/T]$, with $\tau=1$ the time unit.}.

Notice that this protocol is strictly accurate at very low temperatures, 
when `bursts' of plastic activity are well separated in time and the initiation
of new avalanches almost always happen from the weakest site. 
In fact, at any finite temperature it would be possible to wait enough until 
the next initiation simply `happens'. 
This will give the correct physical time scale for time separation among 
avalanches. 
Yet, we don't expect the choice of the first activation to modify qualitatively 
the results.
Within the evolution of each avalanche, the proposed dynamics is run at an 
arbitrary finite $T$ without any approximation.

\begin{figure}[t!]
\includegraphics[width=\columnwidth]{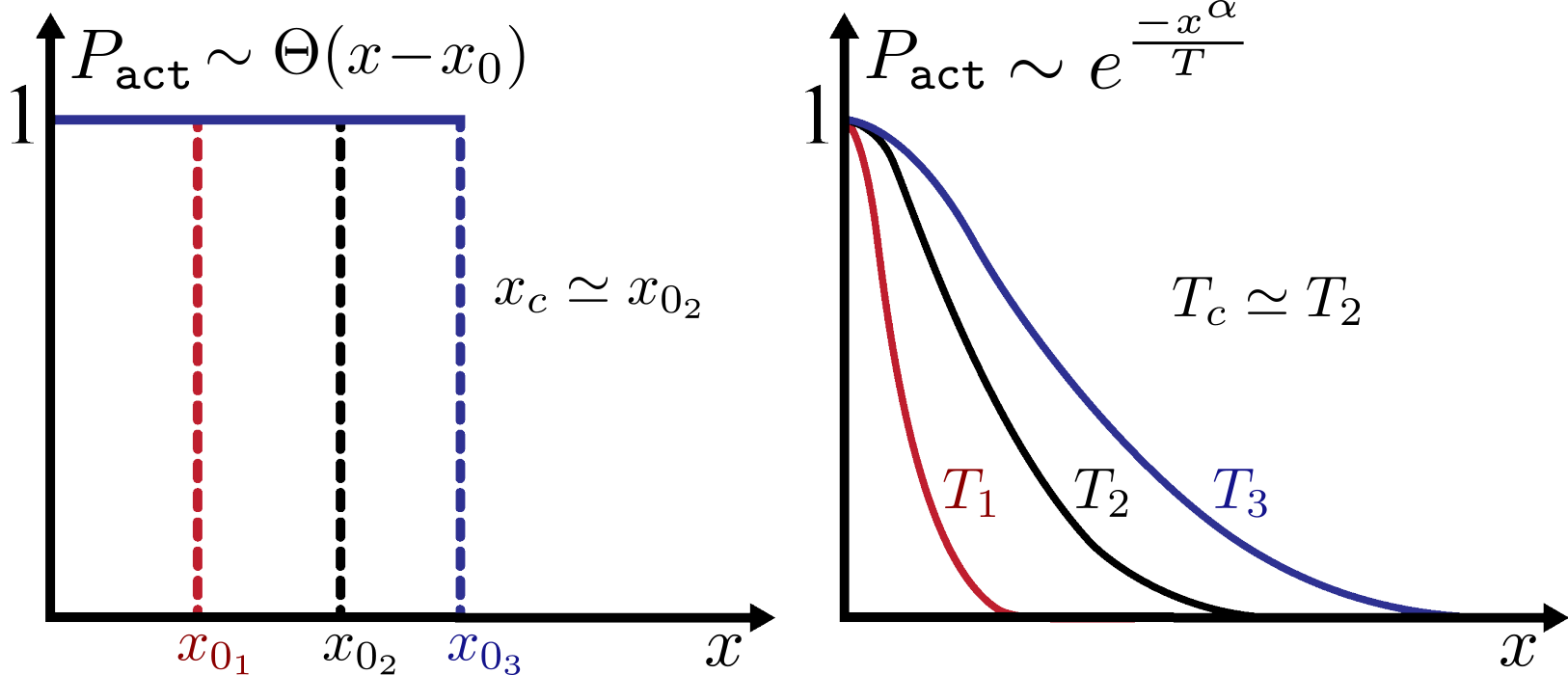}
\caption{\label{fig:schem_protocols} 
\textit{Comparison of the cores of extremal and fully thermal dynamics.}
The schematic curves show the \emph{a priori} probabilities of 
thermal activation of the elastoplastic blocks during the 
avalanche dynamics.
(a) Extremal dynamics, (b) Fully thermal dynamics.
}
\end{figure}

\subsection{Extremal dynamics}
\label{sec:extremaldynamics}

We also implement simulations using an \textit{extremal dynamics} 
protocol~\cite{PaczuskiPRE1996}, \resub{also referred as 
`Arrhenius limit' or `$T=0^+$' protocol in models with vanishing 
temperature~\cite{KoltonPRL2006, KoltonPRB2009, FerreroPRL2017}, 
and recently applied in EPM studies~\cite{OzawaPRL2023, TaheiPRX2023}}.
In \resub{these last implementations} no true temperature is considered.
Instead, a system parameter $x_0$ (assumed to be `small') is defined 
and at each time step every site with $x_i$ smaller than $x_0$ will be 
deterministically activated and added to the ongoing avalanche. 
This is, not only the unstable sites 
($x_i<0$, with $x_i = \min\{(\sigmaY - \sigma_i),(\sigma_i + \sigmaY)\}$).
but also those that are barely stable ($0\leq x_i< x_0$) are activated.
In the absence of sites below $x_0$ the avalanche stops.

The initiation of a new avalanche is implemented by finding $x_{\tt min}$
and its location index, and activating that site, as before. 
So, the difference with the (more realistic) thermal dynamics described above 
is essentially that the activation of a site below threshold ($x_i>0$) 
changes from a probabilistic event to a \textit{deterministic} event, 
described schematically by a step function in Fig.~\ref{fig:schem_protocols}.

When discussing this dynamical protocol, the authors of Refs.~\cite{OzawaPRL2023, TaheiPRX2023} argue that this is a dynamics corresponding to 
vanishing $(T=0^{+})$ temperature. 
As is already visually presented in Fig.~\ref{fig:schem_protocols},
we will argue against this interpretation.
The only step of the dynamics strictly equivalent to a $(T=0^{+})$
is the avalanche initiation.
After that, a dynamics ruled by a finite $x_0$ is an effective 
finite temperature dynamics.
We can think that in the extremal dynamics we replace the standard
Arrhenius activation probability $p_{\tt act} \sim e^{-x^\alpha/k_BT}$ by 
a step function $p_{\tt act}=\Theta(x-x_0)$, with $x_0\equiv x_0(T)$ some explicit 
function of $T$ (for example, the mean value of a normalized probability 
distribution $P(x)=Ce^{-x^\alpha/k_BT}$).

\section{Persistence and dynamical heterogeneities}
\label{sec:dynamicalheterogeneities}

\begin{figure}[t!]
\includegraphics[width=1\columnwidth]{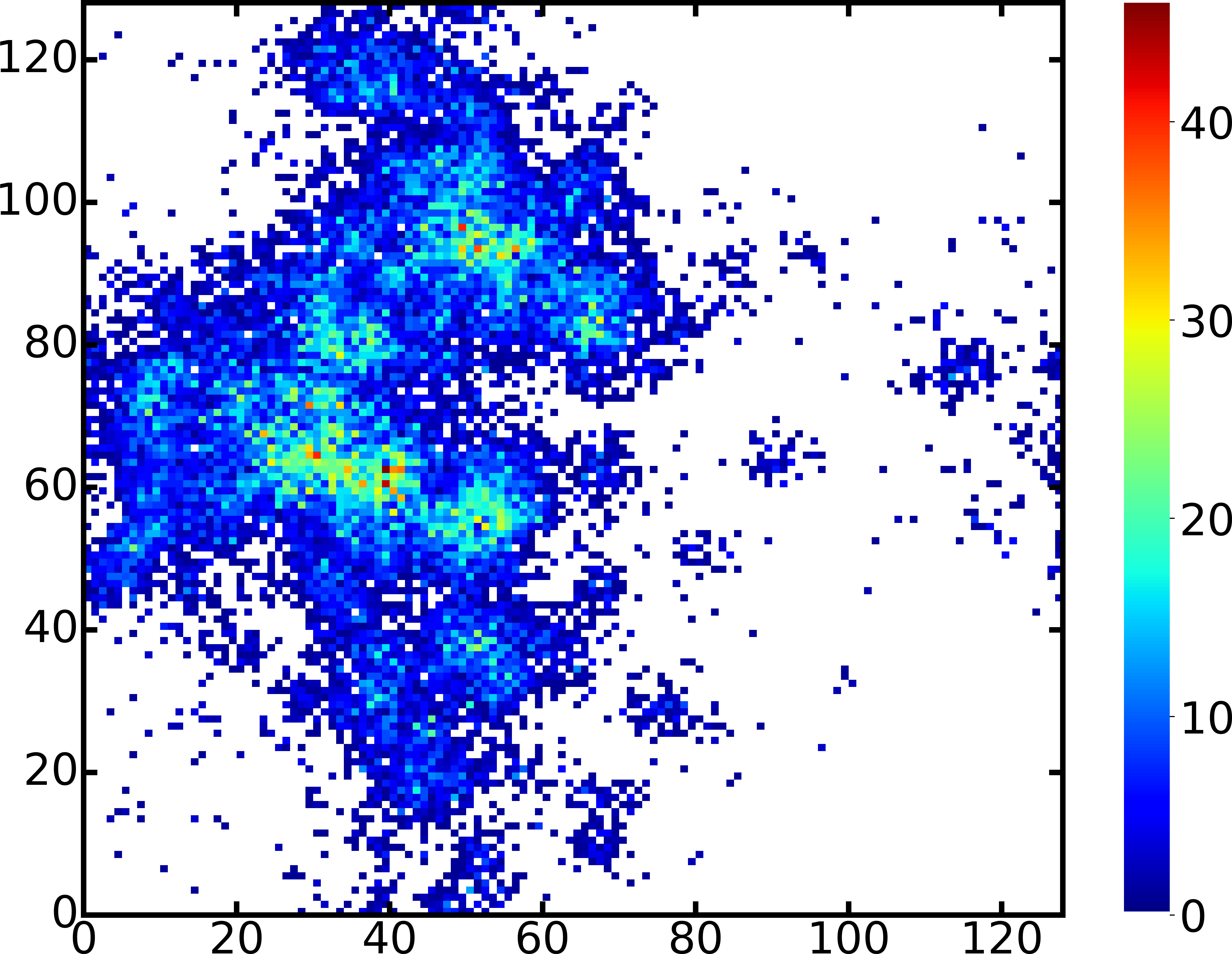}
\caption{\label{fig:map_activity}
\textit{Plastic activity map during a thermal avalanche.} 
The map is constructed by using the number of plastic activations 
$S_{A}(i)$ of each site $i\equiv(x,y)$ during an individual avalanche. 
The color scale on the right indicates the number of activation per site.
The parameters used for this example are $T = 0.0042$ and $L = 128$.
}
\end{figure}

In order to connect with previous literature, we start by studying 
dynamical heterogeneity in our elastoplastic model at finite temperature
using the \textit{thermal protocol} explained in Sec.~\ref{sec:thermaldynamics}.
Figure~\ref{fig:map_activity} shows a map of plastic activity during a 
thermal avalanche in the system.
In this graphical example it is shown already that the 
plastic activity is non-homogeneously distributed, 
but is rather happening at each moment in a heterogeneous manner. 
Similar activity maps, be it for the plastic activity or for the
stress or strain auto-correlation functions, have been argued to be 
a feature of an heterogeneous dynamics akin to the one observed in 
glass-forming materials~\cite{OzawaPRL2023, TaheiPRX2023}.
In~\cite{RodriguezLopezPRM2023} we have shown that the dynamical 
heterogeneity observed in elastoplastic models (without any kind
of quenched disorder or anisotropic fields) is an `itinerant' 
heterogeneity: heterogeneity in space is only temporal and 
disappears when the dynamics is integrated in time.
At each time window observed plastic activity looks very 
heterogeneous in space, but the activity does not occur always
in the same localized places and `moves around' with no 
preferential spots.
Yet, even when different from the dynamical heterogeneity 
of spin glasses~\cite{Ricci-TersenghiPRE2000},
the study of persistence and heterogeneity in 
lattice elastoplastic models at finite temperature 
has been shown to provide interesting analogies with 
the glass transition literature~\cite{OzawaPRL2023, TaheiPRX2023, deGeusPRE2025, GoshSoftmatter2025,KorchinskiPRX2025}.
The `persistence' is usually defined as~\cite{ritort2003glassy, berthier2005, garrahan2011kinetically}

\begin{equation}
p(t)= \frac{1}{L^{2}}\sum_{i} p_{i}(t),
\end{equation}
where $p_{i}(t)=1$ if site $i$ has not yet undergone a plastic event 
up to time $t$, and $p_{i}(t)=0$ if the site has been activated at least once.
In fact, $p(t)$ measures how much correlation is preserved in the system
with respect to a reference configuration taken at the origin of time
(in other words, $t$ measures a time window).
From this observable, one can also define a relaxation 
time $\tau_{\alpha}$, at which the persistence decays below 
specified threshold.
\resub{Following the convention, we define it as $\left<p(\tau_{\alpha})\right>=1/e\simeq 0.36788$.}

\begin{figure}[t!]
\includegraphics[width=0.93\columnwidth]{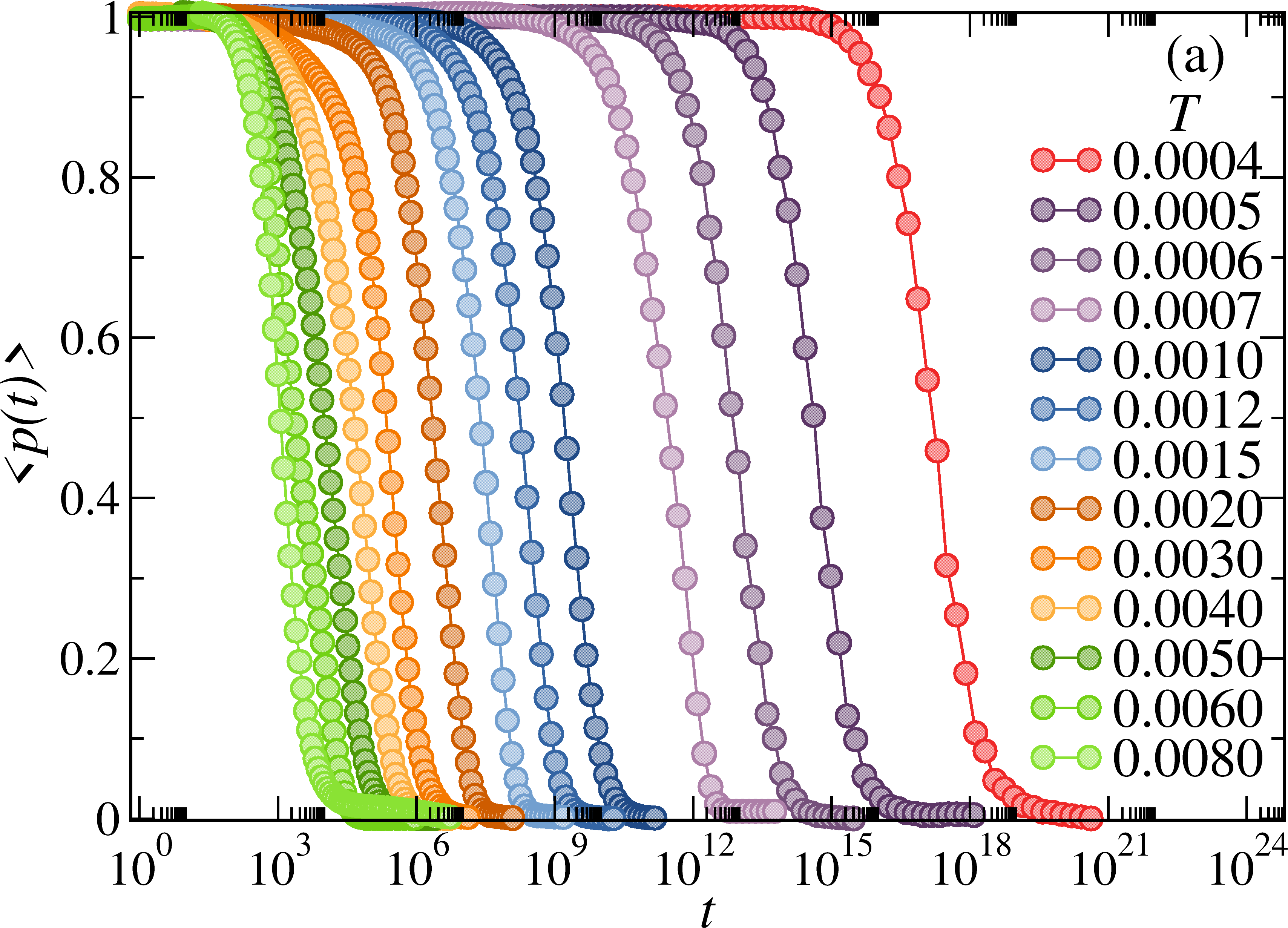}
\includegraphics[width=0.93\columnwidth]{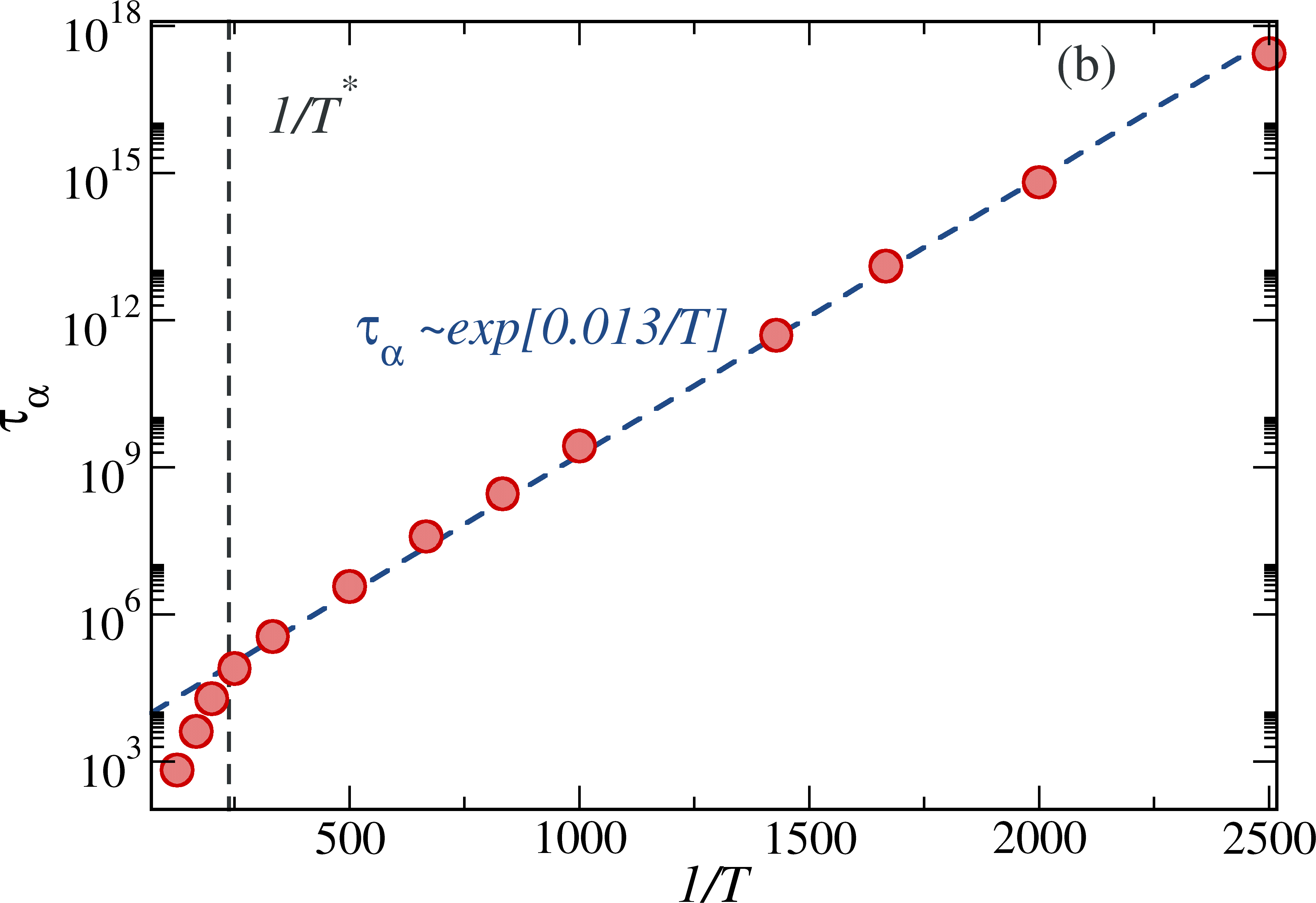}
\includegraphics[width=0.93\columnwidth]{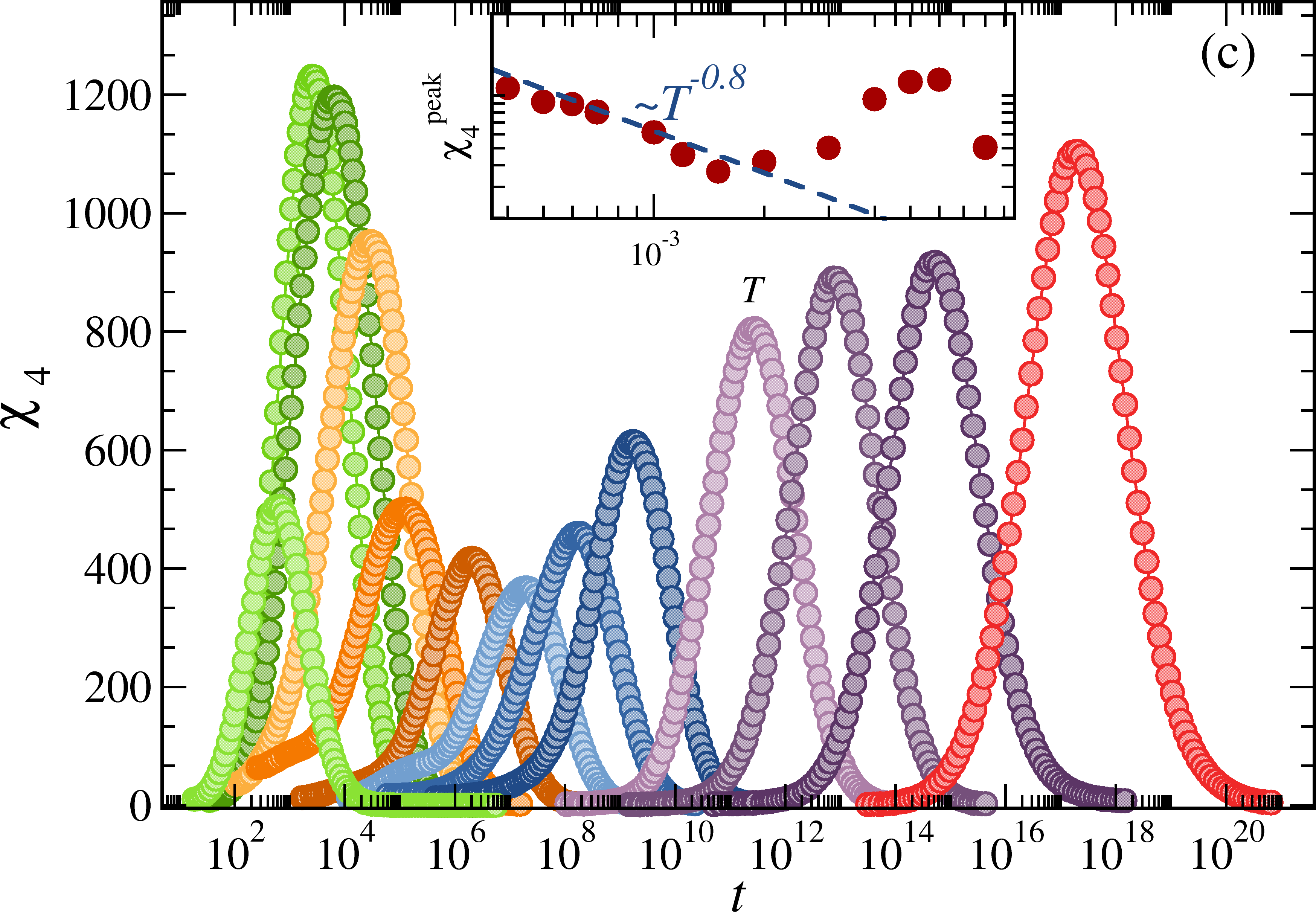}
\caption{\label{fig:persistence_FixSize} 
\textit{Persistence, relaxation time and dynamical susceptibility.}
Panel (a) illustrates the mean persistence curve, $\langle p(t) \rangle$, 
for various finite temperatures at a fixed system size. 
Panel (b) shows the relaxation time $\tau_{\alpha}$ as a function of $1/T$. 
The blue dashed line corresponds to an Arrhenius fit of the form 
$\tau_{\alpha} \propto \exp\left[\frac{0.013}{T}\right]$. 
The gray dashed line indicates the crossover inverse temperature $1/T^*$, 
marking the onset of Arrhenius-like behavior in the curve.
Panel (c) displays $\chi_4$, as defined by Eq.~\ref{eq:chi4} for the 
different temperatures in this analysis 
$T \in [0.0004, 0.008]$
The inset shows $\chi_4^{\textbf{peak}}$ versus $T$.
The blue dashed line corresponds to a power-law fit,
$\chi_4^{\textbf{peak}} \sim T^{-0.8}$.
System size is $L=128$.}
\end{figure}

Figure~\ref{fig:persistence_FixSize}(a) shows mean persistence 
$\left<p(t)\right>$ as a function of time $t$.
Persistence decreases over time at a given temperature $T$ and 
decays earlier with rising temperatures, as expected.
Figure~\ref{fig:persistence_FixSize}(b) shows relaxation time 
$\tau_\alpha$, as a function of inverse temperature $1/T$.
We observe that the relaxation time resulting from the persistence 
curves increases exponentially with $1/T$ for low enough 
temperatures, $\tau_{\alpha}\propto e^{\frac{a}{T}}$, with ${a=0.013}$~\footnote{
As a matter of fact, at low enough temperatures the time delays between 
avalanches of plastic activity are basically dominated by the thermal 
activation of the first site triggered, which by definition of the dynamics 
is $t_{\tt w} = \tau \exp[x_{\tt min}^{3/2}/T]$.
}.
In that sense, our data agrees with recent studies~\cite{OzawaPRL2023,TaheiPRX2023,TakahaPRE2025}.
Yet, let us notice that a crossover to a faster regime occurs at a 
given temperature $T=T^*\simeq 0.0040$ for this data.
\resub{While we} will later relate such a temperature 
with critical aspects of the avalanche dynamics\resub{, 
do to its proximity to an independently defined 
`critical' temperature, one needs to set a warning note 
on the estimation of relaxation times for large temperatures:
\footnote{\resub{The systematic choice of the weakest site for avalanche restarting
is a good approximation at low temperatures but might eventually
be replaced by a true probability tower of activations to have an
accurate time clock (see e.g.~\cite{Russo2026}).
Provided that the waiting times $t_{\tt w}$ can quantitatively 
differ when the total rate of activations is considered, 
the quantitative estimation of $\tau_\alpha$ at large 
temperatures might differ as well.}}.}

We now introduce the so-called susceptibility of the four-point correlation
function~\cite{donati2002,BerthierPhysics2011}, 
$\chi_4(t) = \int d^d{\bf r} G_4(r;t)$,
which in this case can be obtained as the fluctuations of the persistence:
\begin{equation}\label{eq:chi4}
    \chi_{4}(t)= L^{2} \left[\left<p^{2}(t)\right>- \left<p(t)\right>^{2}\right].
\end{equation}
Figure~\ref{fig:persistence_FixSize}(c) displays $\chi_4$ as a function of time 
for the same set of temperatures as in Fig.~\ref{fig:persistence_FixSize}\resub{(a)}.
On one hand, we observe that the position of the peak of $\chi_4(t)$
shifts towards longer times as temperature decreases, this is consistent
with the behavior of $\tau_\alpha$.
Now, at low and decreasing temperatures, as $T \to 0$, also the maximum of the 
susceptibility $\chi^{\tt max}_4(t)$ increases, roughly as a power-law 
$\chi_4^{\tt max} \sim T^{-0.8}$.
This behavior indicates growing dynamical correlation length~\cite{OzawaPRL2023}.
On the other hand, when increasing $T$, a change of regime becomes apparent.
As expected, the position of the $\chi_4(t)$ peak moves to shorter times.
But, while the susceptibility peak height decreases up to $T\simeq0.002$,
as temperature keeps climbing we observe again an increase in the 
$\chi_4^{\tt max}$, not previously discussed.
If the interpretation of the maximum of $\chi_4$ is still correct, 
a correlation length scale is now increasing as $T$ rises, up to 
roughly $T\simeq0.006$ when it drops down again. 
The crossover temperature of the $\tau_\alpha$ behavior, $T^+\simeq0.004$ 
lies somehow in between the deep and the peak ($T\simeq0.002$-$T\simeq0.006$)
of this anomalous regime of $\chi_4(t)$.
The maximum of $\chi_4(t)$ after $T^{*}$ could be signaling a transition or 
dynamical crossover of the system; in particular, an incipient fluidization. 
We now further analyze it by looking at the avalanche statistics. 

As $T$ increases, the peak position of $\chi_4(t)$ shifts to shorter times, as expected. Concerning the peak height, we observe a non-monotonic dependence on $T$: $\chi_4^{\mathrm{peak}}$ first decreases up to $T \simeq 0.002$, then increases up to $T \simeq 0.006$, and finally shows a very weak downturn at our largest simulated temperature $T=0.008$. This behavior is consistent with the onset of a distinct high-$T$ dynamical regime associated with long-lasting, system-spanning avalanches discussed below.

\section{Avalanche statistics for extremal dynamics} 
\label{sec:extremal_avalanches}

\begin{figure}[t!]
\includegraphics[width=0.93\columnwidth]{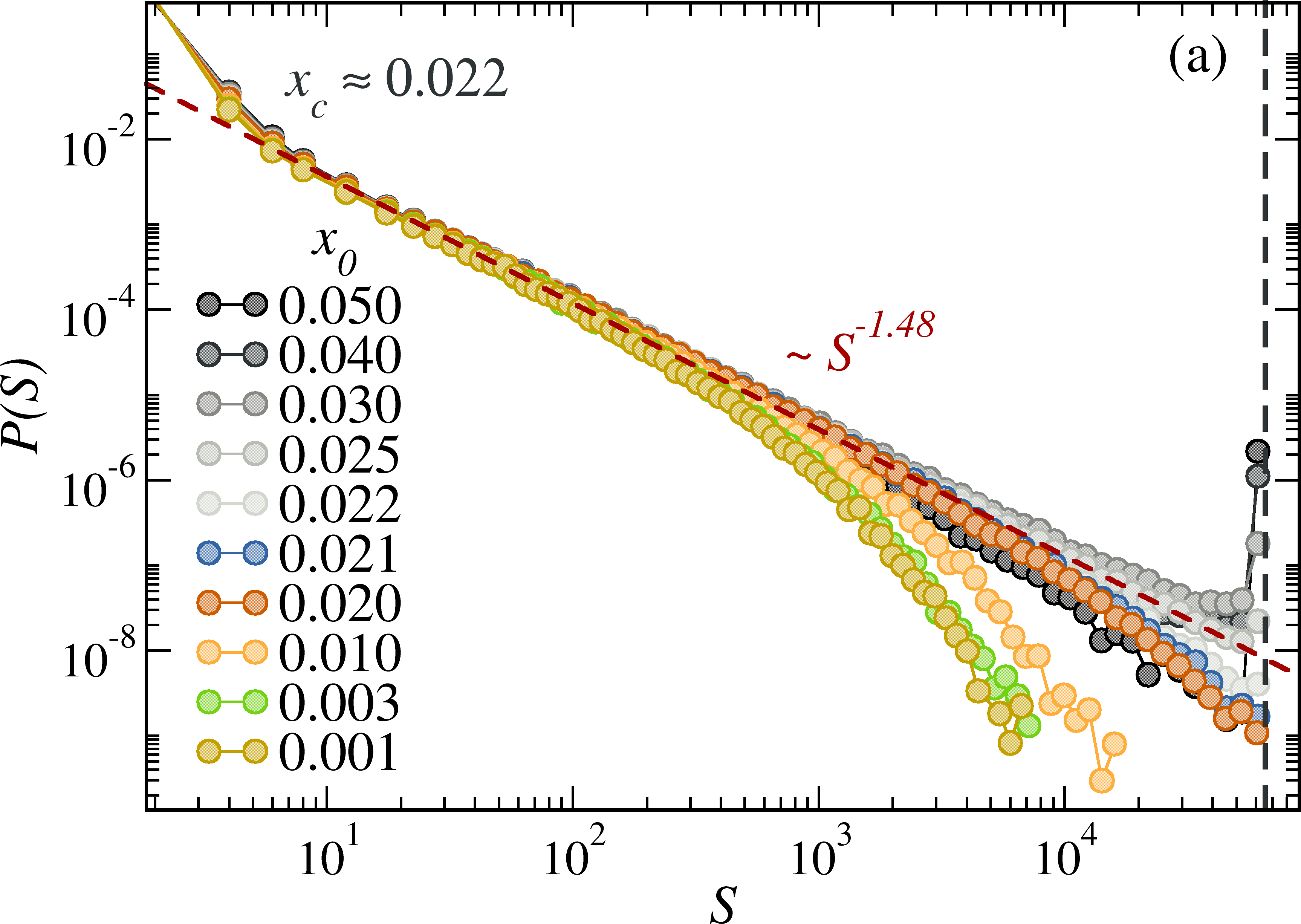}
\includegraphics[width=0.93\columnwidth]{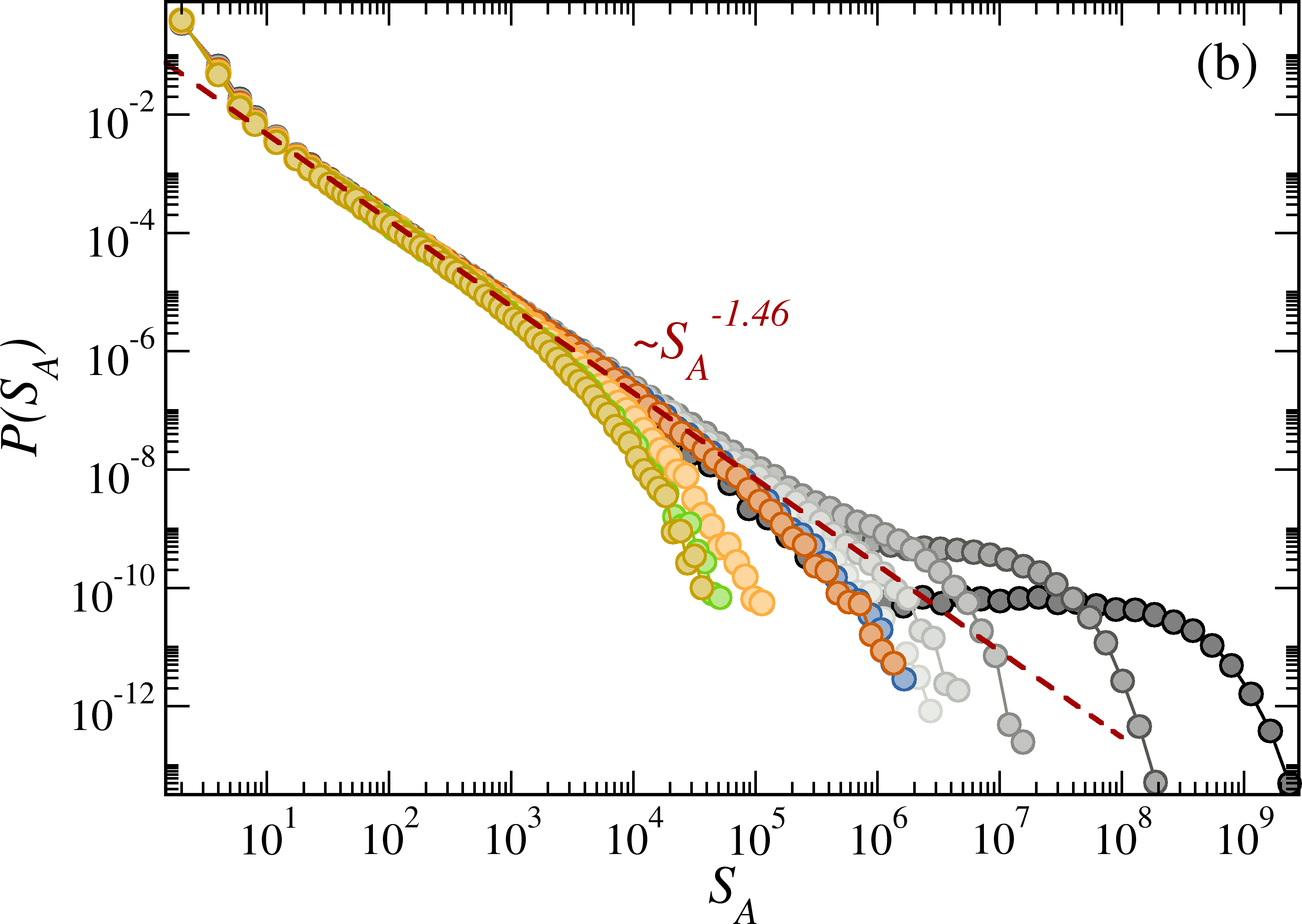}
\includegraphics[width=0.93\columnwidth]{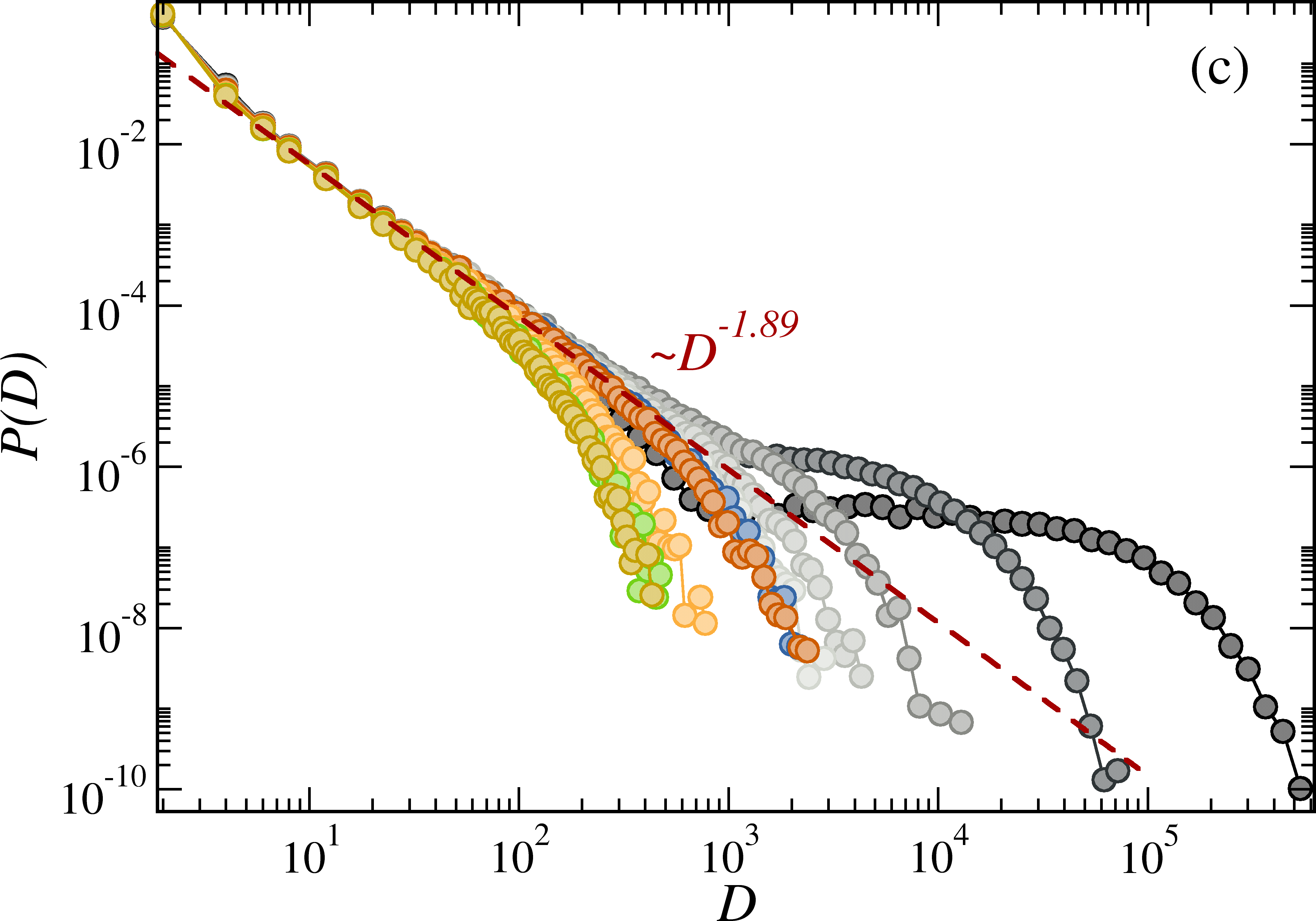}
\caption{\label{fig:extremal_avalanches}
\textit{Extremal dynamics} 
Avalanche size and duration distributions for a 
fixed system size ($L=256$) and $x_0=$[0.001, 0.003, 0.010, 0.020, 0.021, 0.022, 0.025,
0.030, 0.040, 0.050] in the extremal dynamics protocol. 
The critical value is $x_c\approx 0.022$. Curves in grayscale correspond to $x_0>x_c$, 
and the gray dashed line marks the maximum avalanche size, $S_{\text{max}}$.
Panel (a) shows the distribution of avalanche sizes $P(S)$, 
(b) the distribution of the total number of activations $P(S_A)$, 
and (c) the distribution of avalanche durations $P(D)$.}
\end{figure}

To begin the discussion on avalanches, let us start by 
using first the \textit{extremal dynamics} protocol 
(Sec.~\ref{sec:extremaldynamics}).
By running this dynamics in the steady state we collect statistics 
for tens of thousands of avalanches,
and build distributions of avalanche sizes $S$
(measured by the number of sites involved in the avalanche), 
total numbers of activations during an avalanche $S_A$
and avalanche durations $D$
(measured in number of steps of the dynamics elapsed during an avalanche).

Figure~\ref{fig:extremal_avalanches}(a) shows the curves corresponding to $P(S)$ 
for different values of the protocol parameter $x_0$.
From previous work~\cite{TaheiPRX2023, KorchinskiPRX2025}, we expect a power-law
behavior of $P(S)$ vs $S$, $P(S)\sim S^{-\tau}$ for certain critical $x_0$ value.
We observe that $x_c=x_0\simeq 0.022$ can be identified as the value 
that yields the `best' power-law for $P(S)$ vs $S$.
Below $x_0=x_c$, the $P(S)$ curves display a premature cutoff to the power-law, 
which happens earlier (at smaller $S$) for smaller $x_0$.
Above $x_0=x_c$, instead, $P(S)$ displays a sudden overshoot with an increasing
probability accumulation at $S \simeq N$ as $x_0$ increases.
This behavior at $x_0 >x_c$ hasn't been reported in previous works.
The power-law regime for $x_0 \simeq x_c$ displays $P(S)\sim S^{-\tau}$ with 
$\tau=1.48$, a value for the $\tau$ exponent consistent with the one 
observed by Korchinski et al.~\cite{KorchinskiPRX2025} ($\tau\simeq1.5$).

Figure~\ref{fig:extremal_avalanches}(b) shows $P(S_A)$ for the same set of $x_0$ 
values.
$P(S_A)$ curves corresponding to $x_0<x_c$ are very similar to the $P(S)$ curves.
At $x_0=x_c$ again we find the `best' power-law, although the exponent slightly 
differs from the one of $P(S)$, $P(S_A)\propto S_A^{-1.46}$.
But now, above $x_c$, $P(S_A)$ display a `shoulder' at large $S_A$.
From the correspondence between the behaviors of $P(S_A)$ and $P(S)$,
we understand that the avalanches contributing to these shoulders are
system-filling avalanches, which are eventually exhausted after different 
amounts of activations (necessarily multiple in several sites, if we 
compare for example a cutoff in the order of $S_A \sim 10^9$ for 
$x_0=0.05$ with the system size $N\sim 6\times10^5$).
The shoulder or plateau in $P(S_A)$ is depleted by an exponential cutoff; 
in fact, all the regime after the power-law could be simply an exponential 
distribution, probably rooted in a Poissonian stochastic process for the 
dynamics of extinction of a system-filling avalanche once it takes place.

\begin{figure}[t!]
\includegraphics[width=0.9\columnwidth]{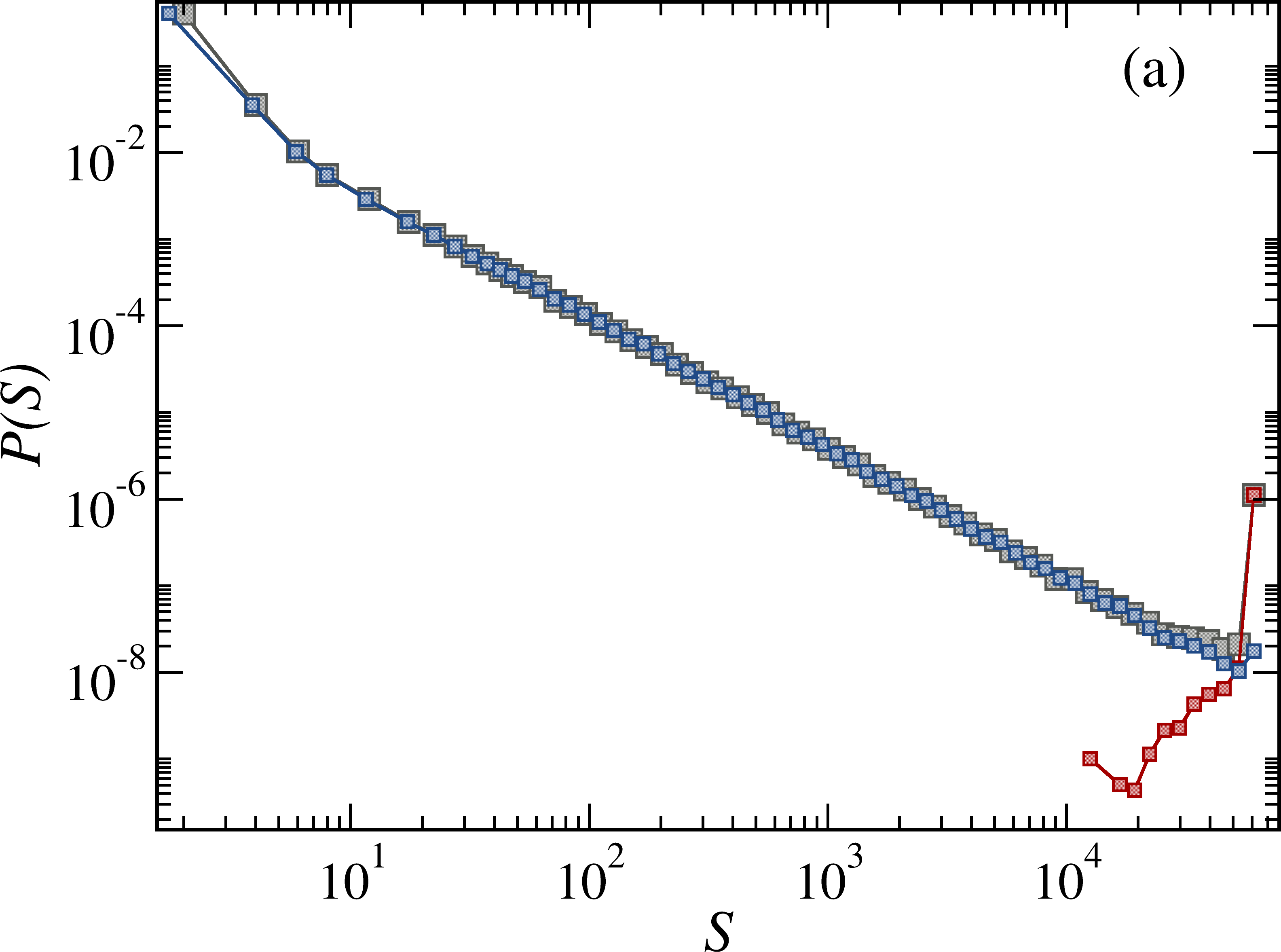}
\includegraphics[width=0.9\columnwidth]{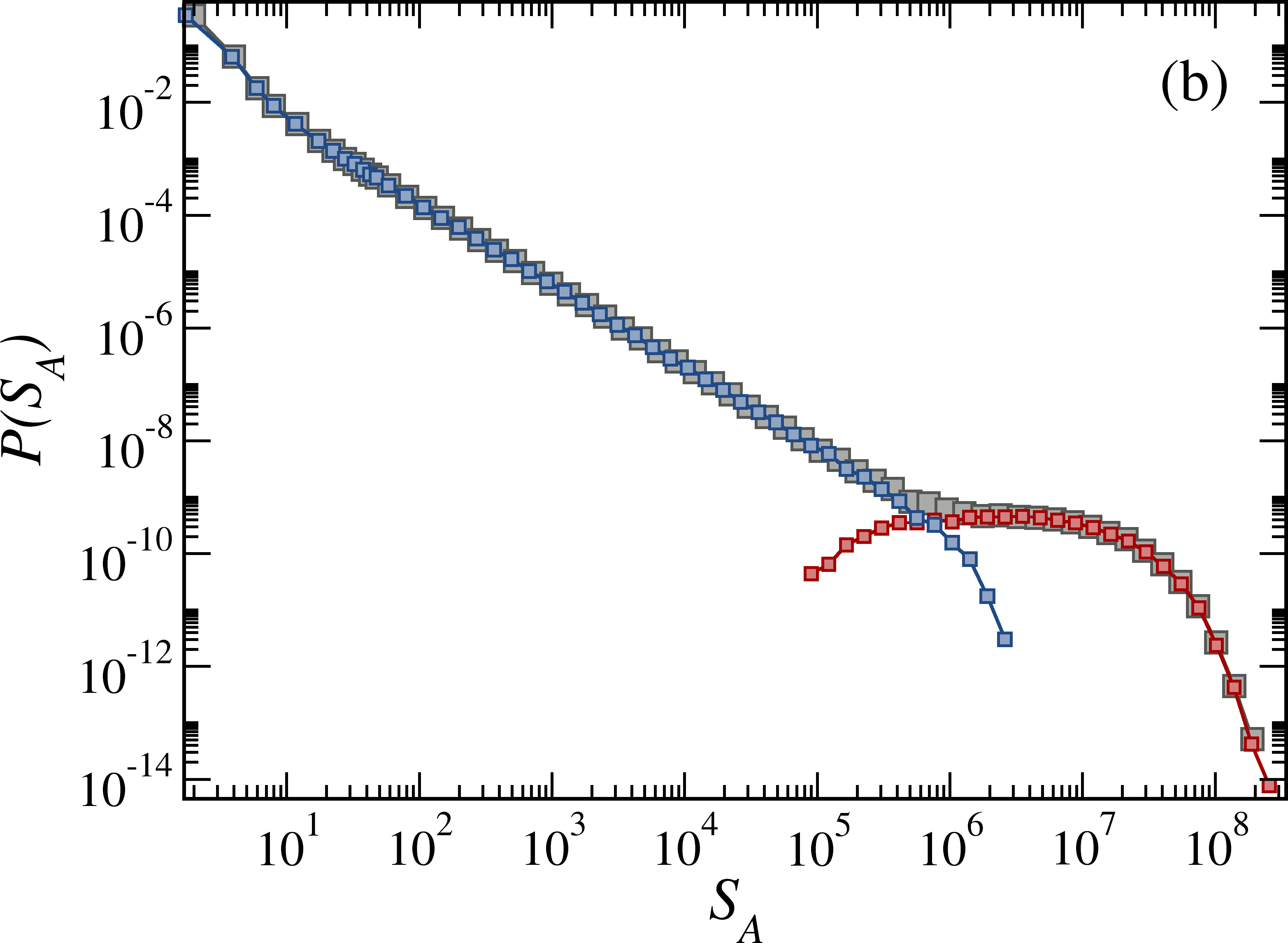}
\caption{\label{fig:p_s_sa_split_size256_extremal}{
Partial contributions to the avalanche size distributions $P(S)$ and $P(S_A)$
in the extremal dynamics for $L=256$ and $x_0=0.04$.
Red circles represent the distribution of avalanches with durations 
below a threshold $D^{*}\simeq900$, 
while blue circles denote the distribution
of avalanches exceeding $D^{*}$. 
Gray squares illustrate the distribution encompassing all avalanches. 
}
}
\end{figure}

The distributions $P(D)$ of avalanche durations $D$ are shown in 
Fig.~\ref{fig:extremal_avalanches}(c).
They look very similar to the distributions of $S_A$, yet again
the exponent that one can estimate for the power-law region is
different from the previous ones: $P(D)\propto D^{-1.89}$. 
The characteristic `shoulders' or `bumps' displayed by $P(D)$,
similarly to what happens in $P(S_A)$, that become more and more 
important as $x_0$ grows beyond $x_c$, tell us about a change of 
dynamical regime.
The avalanches contributing to the bump are avalanches that have 
already saturated the system size at some point of their lifetime, and
ceased to be ruled by scale-free critical behavior. 
Instead, when this happens, the dynamics turns into a dynamics
of survival probabilities.
We can always ask what is the probability for at least one site 
to be activated in the next time step (and the avalanche
remain active). 
If we depart from the point of large number of active sites ($\sim N$)
and large threshold depletion ($x_0>x_c$), we could 
think that such probability is high and therefore the probability for 
the avalanche to stop is low.
Even when that survival probability is not a constant, because the
avalanche keeps evolving, it seems from the data that in fact
runaway avalanches describe a Poissonian process for their lifetime
(both in Fig.\ref{fig:extremal_avalanches}c and 
Fig.\ref{fig:p_s_sa_split_size256_extremal}b
one distinguishes a nearly flat plateau before the fast cutoff and this 
is well described by an exponential).
We could interpret that the supercritical value of $x_0$ is controlling 
that survival probability (the \textit{a priori} probability of activating 
at least one of the $N$ sites per time unit) and that it fluctuates around 
a finite value. 
That value controls the tail of $P(D)$ and $P(S_A)$.

A similar shoulder in the avalanche statistics distributions of sizes and durations
occurs in inertial systems~\cite{SalernoPRL2012}, when it becomes harder and harder 
to stop an avalanche, but it typically ends up developing a characteristic peak\cite{KarimiPhysRevE2017}.
Interestingly, such a `bump' in the distributions has also been observed 
and attributed to `supercritical' avalanches in depinning 
recently~\cite{LaursonPRL2024}.
A related phenomenology has been reported in other nonequilibrium 
disordered systems exhibiting spreading dynamics, such as 
epidemic-like activity models.
In particular, in cooperative contagion models, overcoming a 
critical control parameter $p_c$ results in infected nodes 
forming giant (percolating) clusters~\cite{NaturePhysicsCai2015},
displaying a similar self-reinforcement dynamics as the one of 
our avalanches above $x_c$.

To better understand the effect of long-duration avalanches 
we divided the events into two groups based on their duration $D$, 
using a threshold value $D^{*}$ as a criterion. 
The threshold is chosen by visually inspecting when $D^{*}$ has a kink
or change of regime.
Avalanche events were separated into `short' ($D \leq D^{*}$) and 
`long' ($D > D^{*}$) duration subsets.
We then computed partial probability distributions for $P(S)$ and $ P(S_A)$, 
with each subset, normalized by its respective fraction of avalanches.
Figure~\ref{fig:p_s_sa_split_size256_extremal} shows the results 
of this analysis for a system of size $L=256$, running the extremal 
dynamics with $x_0 = 0.04$, We visually estimate the value of $D^*$, 
examine $P(D)$, and define $D^{*}$ as the point of inflection or critical
threshold of $P(D)$, which in this instance is approximately $\simeq900$.
Two distinct and related effects are observed on the right side of the 
distributions: 
a prominent \textit{peak} in $P(S)$ and a \textit{plateau} in $P(S_A)$.
The separation in two contributions clearly shows that the responsible of 
such features are the long-duration avalanches, which tend to involve all, 
or almost all, available sites in the system ---which explains the emergence 
of the pronounced peak in $P(S)$.
It also explains the shoulder in $P(S_A)$, because system-filling avalanches 
start to reinforce themselves and growing beyond $S_A\sim N$ means multiple 
activations for each site on average; long-last duration is directly correlated 
with repeated activations per site.
On the other hand, in the subset of avalanches with $D \leq D^{*}$, the sharp 
peak in $P(S)$ disappears and the distribution simply-shows a scale-free behavior. 
A similar effect is observed in $P(S_A)$, where the plateau vanishes and is replaced 
by an earlier cutoff.
Essentially, the subset of avalanches with durations below $D^*$ behave as 
classical overdamped avalanches, with distributions that exhibit a 
well-defined power-law regime, eventually followed by fast cutoff. 

\begin{figure}[t!]
\includegraphics[width=0.93\columnwidth]{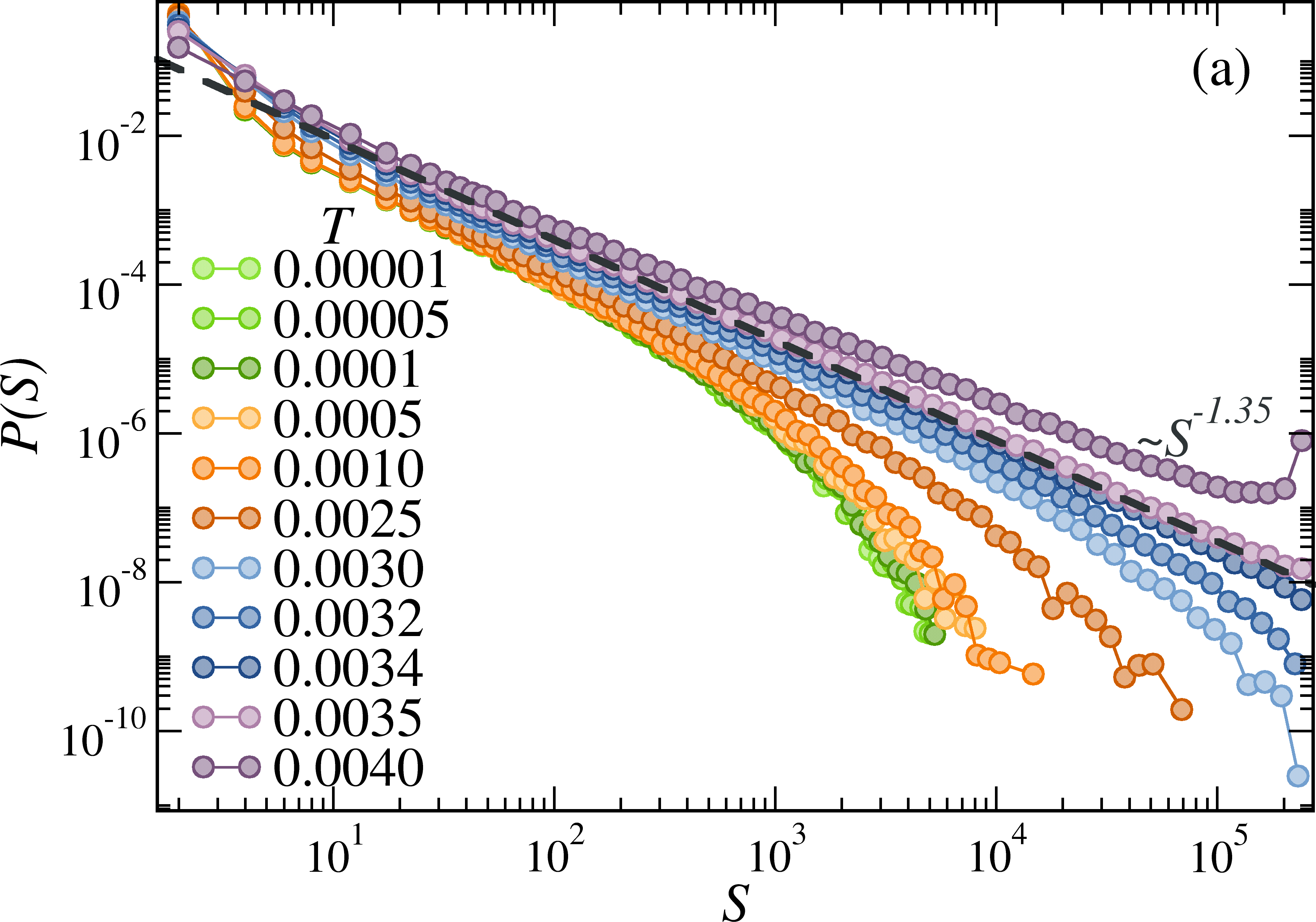}
\includegraphics[width=0.93\columnwidth]{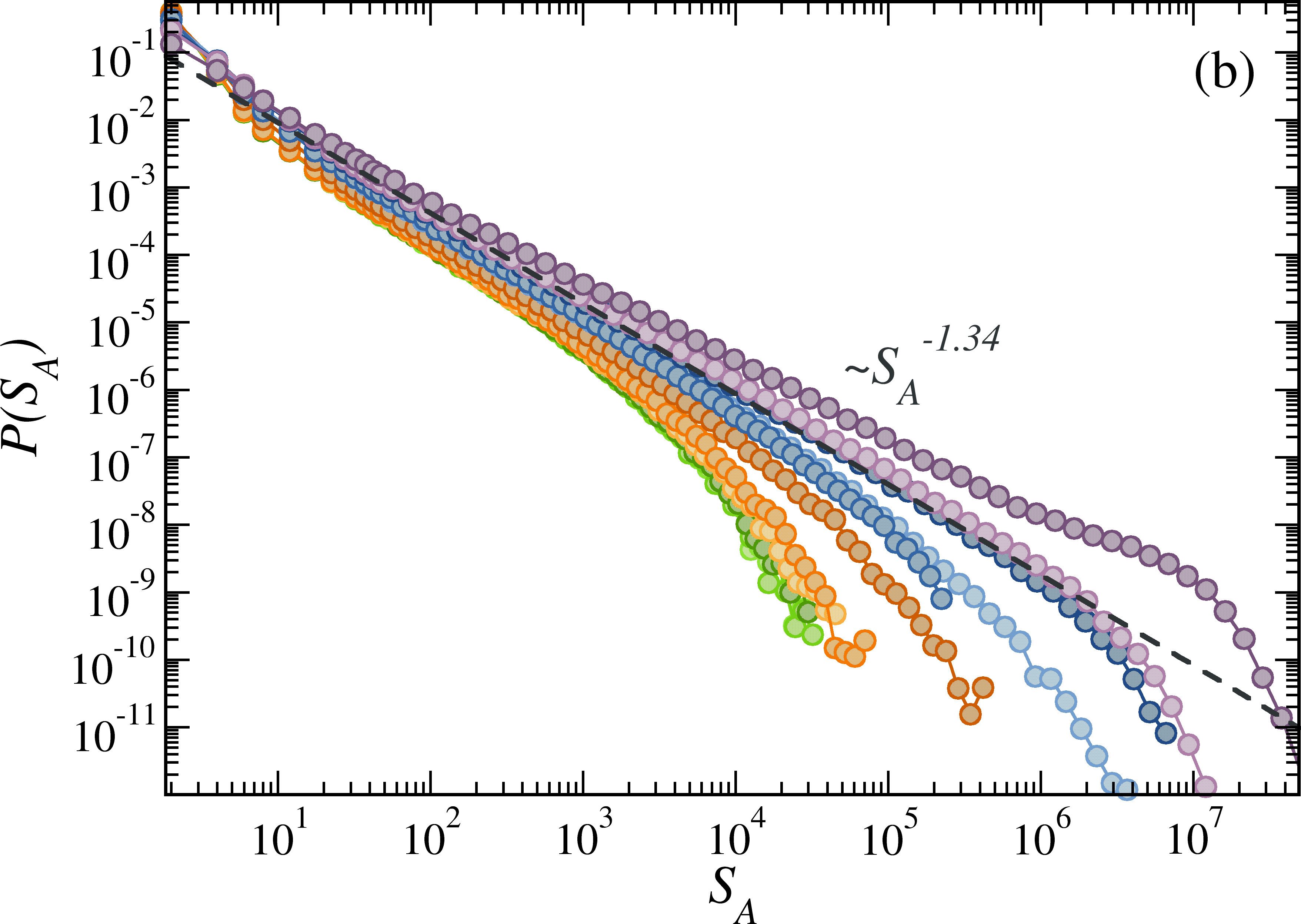}
\includegraphics[width=0.93\columnwidth]{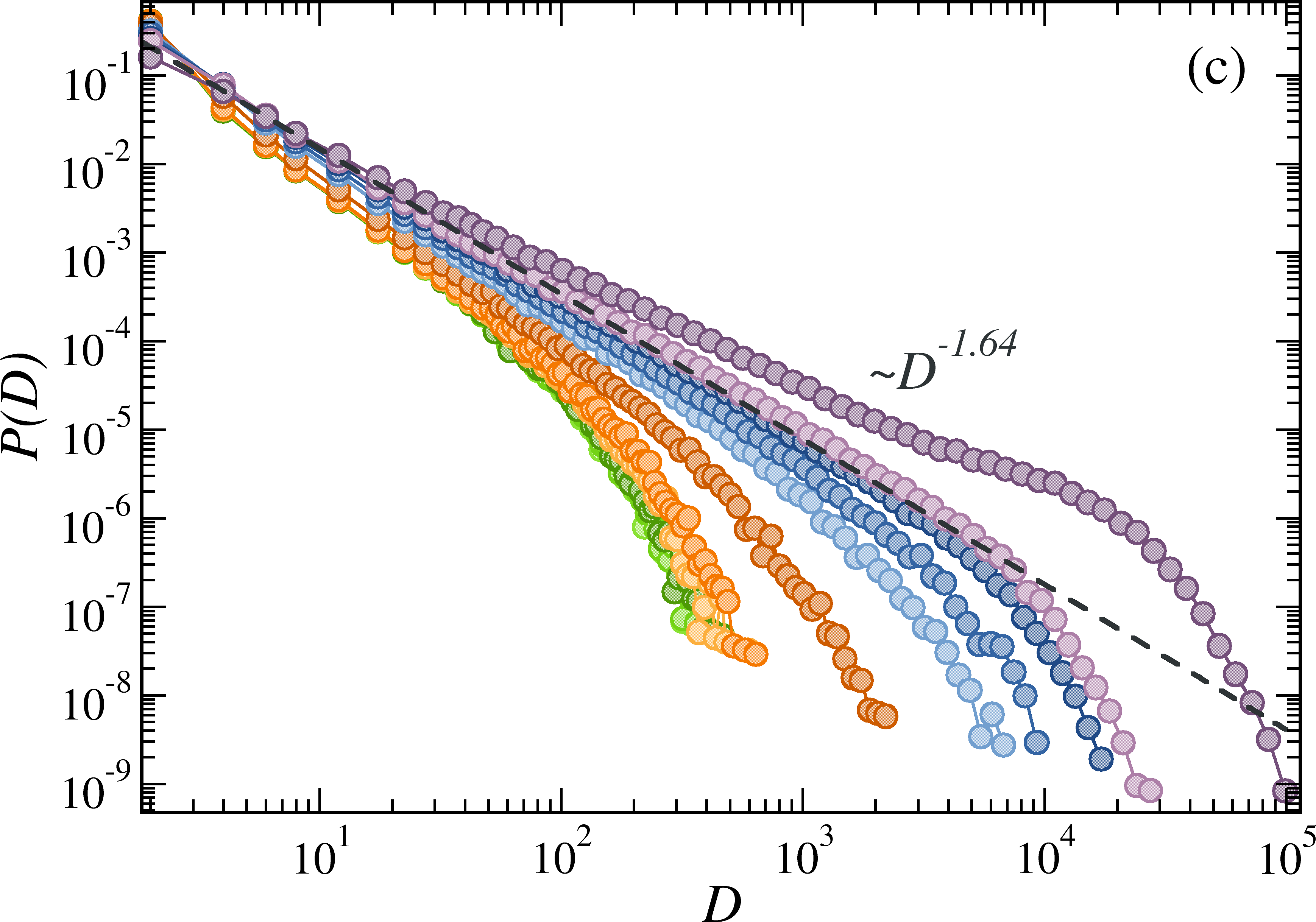}
\caption{\label{fig:thermal_avalanches}{
Avalanche size and duration distributions for a fixed system size ($L=512$) at different temperatures 
$T$ = [0.00001, 0.00005, 0.0001, 0.0005, 0.0010, 0.0025, 0.0030, 0.0032, 0.0034, 0.0035, 0.0040]:
(a) Probability distribution of the number of sites involved in an avalanche, $P(S)$.
(b) Probability distribution of the total number of activations during an avalanche, $P(S_A)$.
(c) Probability distribution of avalanche durations, $P(D)$.
Dashed lines indicate the power-law behavior at the critical temperature $T_c$:
$P(S)\propto S^{-1.35}$, $P(S_A)\propto S_A^{-1.34}$, and $P(D)\propto D^{-1.64}$.
}}
\end{figure}

\section{Avalanche statistics at controlled temperatures}
\label{sec:avalanchedistributions} 

We now come back to the fully thermal dynamics described in 
Sec.~\ref{sec:thermaldynamics}.
We construct the distribution curves for the number of 
sites involved in an avalanche, $P(S)$, the total number of
activations during an avalanche, $P(S_A)$, and the avalanche 
duration, $P(D)$, for a system of size $L = 512$ and several 
different temperatures in the range $T \in [0.00001, 0.004]$. 

Looking to the $P(S)$ curves in Fig.~\ref{fig:thermal_avalanches}(a), we 
first observe at low temperatures a power-law behavior interrupted by 
a cutoff that progressively shifts to the right as we increase the 
temperature.  
Right as in the case of the extremal dynamics, an increase of $T$ beyond
a certain threshold value gives rise to the occurrence of system-filling
avalanches. 
Beyond this critical temperature $T=T_c$, avalanches tend to occupy 
the entire system, and the avalanche dynamics is self-sustained.
When we overcome $T_c$, at large enough temperatures, the cutoff of 
the $P(S)$ distribution is replaced by a formation of a `peak'.
In fact, at the critical value -identified at $T_c\simeq0.0035$ for 
this system size-, the power-law survives essentially with no 
cutoff until $S=L^d$.
Identifying the $L$-dependent critical temperature is crucial, since for 
$T \geq T_{c}$, the duration of avalanches increases significantly. 
The overall behavior of $P(S)$ in this controlled temperature 
protocol is very similar to the one shown in the previous section~\ref{sec:extremal_avalanches}, 
which allows us to argue that the `extremal dynamics' presented 
in~\cite{OzawaPRL2023,TaheiPRX2023} is, in fact, an \textit{effectively
finite–temperature dynamics} rather than a $T = 0^{+}$ one.
One difference that we observe with respect to the extremal dynamics 
data is that, now, the power-law region seems to be smoothly changing 
(varying its exponent $\tau$) as we approach $T_c$ from below.
We will come back to this discussion.

Figure~\ref{fig:thermal_avalanches}(b) shows $P(S_A)$ distributions
at different temperatures.
They follow $P(S_A) \propto S_A^{-1.34}$ at the critical temperature,
with an exponential cutoff. 
For $T > T_c$, a shoulder emerges in the distribution, right as in the
extremal dynamics for $x>x_c$. 
Again, the difference is that now the $\tau$ exponent is not preserved.
Power-laws are steeper (larger exponent) at low temperatures and the 
exponent decreases as the temperature rises.

Finally, in agreement with the above results, the avalanche duration 
distribution $P(D)$ (Fig.~\ref{fig:thermal_avalanches}(c)) exhibits 
a change of characteristics when the temperature goes above $T_c$, 
with a shoulder developing at the largest shown temperature. 
A power-law regime can be identified for every temperature, being 
$P(D)\propto D^{-1.64}$ at $T=T_c$.
In the case of the duration $D$ the shift in the power-law exponent 
as we sweep through the range of temperatures is the most notorious 
one, going from $\tau\simeq 2.0 $ at $T=10^{-5}$ to 
$\tau\simeq 1.35$ at $T=0.004$
Again, it is possible to discriminate contributions to the 
avalanche size distributions based on avalanche duration.
The picture is similar to the one observed in the extremal dynamics
case, see Fig.~S8 in the Supp. Mat.~\cite{SM}.

\subsection{Discussion on thermal avalanche dynamics}

A partial conclusion from this comparison of avalanche statistics 
under the different dynamical protocols is that, in the end,  
the \textit{extremal dynamics} produces a quite a similar picture 
than the one of the \textit{thermal dynamics}, which is a finite 
temperature protocol in a more classical sense.
So that, the \textit{vanishing temperature} ($T=0^+$) 
interpretation of the extremal protocol does not really stand, 
at least not from the moment that a $x_0\neq 0$ is used. 
As a contrasting example, a true $T=0^+$ dynamics can be found applied 
in~\cite{KoltonPRB2009, FerreroPRL2017} to models of elastic interfaces.

Still, it should be said that the extremal dynamics' results come out
somehow `cleaner' and easier to interpret, while the true thermal 
dynamics is more involved.
In contrast to what is observed in Fig.~\ref{fig:extremal_avalanches}, 
where cutoff of the distributions $P(S), P(S_A), P(D)$ shifts with $x_0$ 
while the exponents of the power-law remains unchanged, in 
Fig.~\ref{fig:thermal_avalanches} we observe that the 
power-law regime itself is temperature-dependent. 
Simulation results in system of different sizes ranging from $L=64$ to $L=2048$
indicate that, as the temperature increases, the exponent $\tau$ governing
the power-law of $P(S)$ decreases transitioning from roughly $\tau\simeq1.5$ 
to $\tau\simeq1.1$.
The changes in the $\tau$ exponent in the range of temperatures studied
is less evident for small system sizes, 
where curves look similar to the ones produced by the extremal dynamics, 
with an exponent $\tau\simeq1.5$ fairly preserved (see Fig.~S1 in 
Supp. Mat.~\cite{SM} for $L=64$).

Similarly, the distributions $P(S_{A})$ and $P(D)$ 
exhibit this temperature-dependent behavior.
In Fig.~\ref{fig:thermal_avalanches}(b-c) we see power-law exponents
changing from $\sim 1.45$ to $\sim 1.19$ for $P(S_{A})$
and from $\sim 2.0$ to $\sim 1.35$ for $P(D)$, respectively.
Again, this spreading behavior in the distribution curves (due to the 
change of power-law exponents with $T$) is more and more evident
for larger system sizes, $L = 512, \ldots, 2048$;
this is, when $T_c(L)$ decreases and overlaps with 
the studied range of temperatures \resub{(see Sec.~\ref{sec:criticaltemperature})}. 
The apparent drift of $\tau$, therefore, reflects the increasing 
contribution of long-lived, merging avalanches as the system 
approaches $T_c(L)$ from below.
In other words, beyond the mere formation of individual events, one must 
also consider the possibility that these avalanches may coalesce.
This effect also is discussed in~\cite{KorchinskiPRX2025}, where 
power-law predictions rooted in the critical avalanche statistics
are only valid up to a certain system size (or the ability to 
separate avalanches).
Large systems probe the dynamics very close to the stability limit
(there's always a very small $x$ available) and therefore large 
avalanches are formed frequently through coalescence.
In the terms of~\cite{KorchinskiPRX2025}, these belong to a distinct 
percolation universality class. 
The observed variation of the exponent $\tau$ could be therefore 
related to coalescence effects, \resub{and we are inclined toward
this interpretation}.

\resub{While the `merging' or co-occurrence of otherwise uncorrelated 
plastic activity in a single avalanche} could provide a plausible 
\resub{explanation} to the observed trends, it cannot be unambiguously 
\resub{disentangled} from finite-size effects.
%
In fact, another angle of explanation of a stronger effect of temperature 
for larger systems is that: as the system size grows, the probability 
of sustaining avalanches by triggering new plastic events in the system, 
also increases. 
\resub{Since the co-occurrence happens more easily for larger and 
larger system sizes, it can be considered also to be a finite-size effect, 
yet not in the limiting sense (as if the effect should disappear in the 
thermodynamic limit) but in the facilitation sense instead. 
}
This effect has also been reported by Ikeda's group in~\cite{TakahaPRE2025} 
(a study based on molecular dynamics simulations).
In particular, they observe that ``in larger systems, the 
rearrangements occur more frequently'', and therefore
they need to ``use smaller time intervals to isolate individual 
rearrangement events'', which of course requires ``huge numerical costs''.

\resub{Notice in particular that, if we step at $T_c(L)$ 
(temperature that decreases with increasing system size after $L=128$),
we still see a wider and wider distribution of avalanche sizes $P(S)$ 
as $L$ is increased, with an effective $\tau$ exponent decreasing 
from $\sim 1.5$ to $\sim 1.1$ (see Supp. Mat.~\cite{SM}).
In this case, we are lowering the temperature and consequently the 
mean plastic activity as we increase the system size 
(following the premise of the $T_c$ definition), 
yet the growing system size manages to favor avalanche coalescence 
and the same trend of variation of exponents in $P(S)$ as in 
the case of a fixed temperature (Fig.~\ref{fig:thermal_avalanches}).
}

\resub{Finally, let us clarify that the mechanism of avalanche merging 
or co-occurence does not imply an increased plastic activity.}
The mean instantaneous plastic activity (density of simultaneous active sites) 
remains rather small even at large temperatures. 
Furthermore, it decreases with system size and seems to stabilize 
at large $L$ (see Table 1 in the Supp. Mat.~\cite{SM}).
\resub{Concurrently, this happens to be} important in order to weight the 
relevance of EPM results, since plastic event screening can be relevant at 
high activity densities and elastic response becomes 
anomalous~\cite{LemaitrePRE2021, KumarEPL2024, Tarjus_arXiv2026}.

\subsection{Critical temperature $T_{c}$ for different system sizes}
\label{sec:criticaltemperature}

\begin{figure}[t!]
\includegraphics[width=0.9\columnwidth]{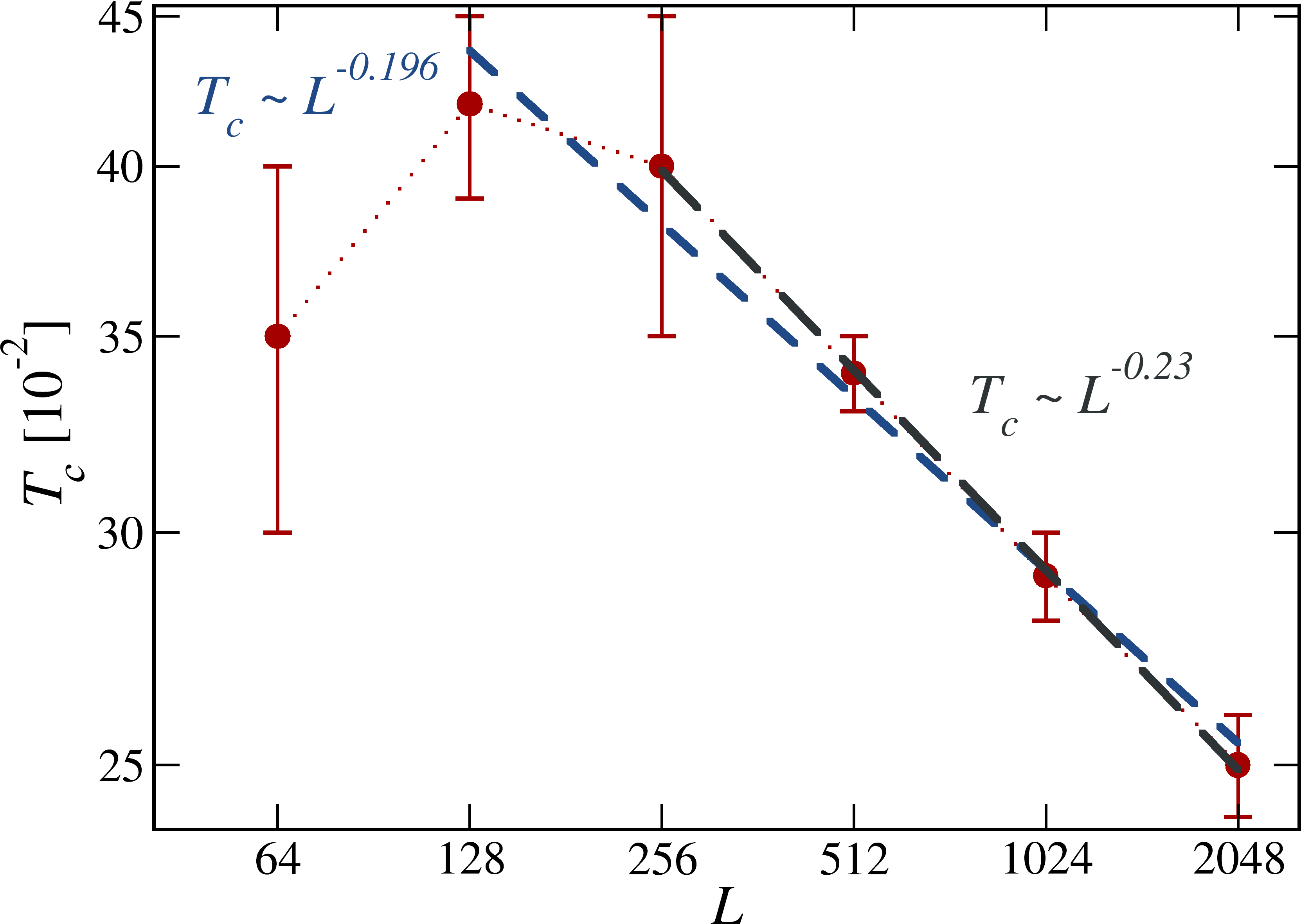}
\caption{
\textit{Critical temperature $T_{c}$ versus Size $L$ in the stationary state}. 
The blue dashed line represents the following fit $T_{c}\propto L^{-0.196}$, and 
the gray dotdash line follows $T_{c}\propto L^{-0.23}$.
}
\label{fig:Tc_sizediff}
\end{figure}

In a similar way as has been done in Sec.~\ref{sec:extremal_avalanches} 
to define $x_c$, in order to define a critical temperature $T_c$
we look at the distributions of $P(S)$ at different temperatures,
e.g., Fig.~\ref{fig:thermal_avalanches} for $L=512$ and similarly 
for other system sizes in the Supp. Mat.~\cite{SM}.
We find that, as we increase the temperature, we reach a value where 
the $P(S)$ cutoff essentially disappears, and, at larger $T$
the power-law not only lacks a cutoff but also develops a `peak'.
The transition temperature at which this occurs can be defined for
each system size and called $T_c(L)$. 
We obtain $T_c(L) \simeq [0.0035, 0.0042, 0.0040, 0.0034, 0.0029, 0.0025]$ for
$L= [64, 128, 256, 512, 1024, 2048]$ (see Table 1 in~\cite{SM}).

Figure~\ref{fig:Tc_sizediff} shows the critical temperature $T_c(L)$ 
as a function of the system size $L$. 
After an initial increase (or plateau, considering the error 
bars\footnote{The uncertainty is computed from the separation between enclosing 
curves of temperatures below and above $T_c$ for which the $P(S)$ behavior clearly 
deviates from a power law.}), we observe that the critical temperature
\textit{decreases} with $L$.
The decay is consistent with a power law $T_{c}\propto L^{-\eta}$,
with the value of $\eta$ around $\eta \simeq 0.23$, 
so $T_{c}\propto N^{-0.115}$.
Of course, being $\eta$ a small exponent, a logarithmic decrease cannot 
be ruled out within error bars, yet the power-law decrease seems to
be the case.
The decay with system size is slow, but this is saying nevertheless 
that in the thermodynamic limit ($N\to \infty$) the critical temperature 
tends to zero ($T_c \to 0$),
This suggests that in an infinite system even arbitrarily small 
temperatures may trigger system-spanning avalanches.
As the size increases, the probability of activating new sites while 
an avalanche is running also increases.
At some point there's a temperature at which on average 
the probability of activating new sites somewhere in the system 
(the `frequency' of plastic events of Takaha~\textit{et al.}\cite{TakahaPRE2025}) 
becomes very high, close to one, therefore sustaining runaway avalanches.
Consequently, increasing the system size reduces the critical temperature.

One should be cautious to interpret $T_c(L)$.
Perhaps one should think it more as a finite-size instability scale 
rather than a sharp thermodynamic phase transition;
yet it's effect points towards a solid-to-fluid change of 
behavior nonetheless.
\resub{Let us dig a bit more into this:
Something similar} happens for the extremal 
dynamics studies, larger system sizes result in smaller 
values of $x_c$. 
In~\cite{OzawaPRL2023, TaheiPRX2023} this has been put as:
``in order for the extremal dynamics (or  $T=0^+$ assumption) 
to be valid, one needs to limit the system size to 
$N\ll N_T \sim T^{-1/\delta}$''.
\resub{The rationale of the low-temperature behavior, 
e.g. the $\chi_4^{\tt peak}$ vs $T$ scaling, 
is that there is a temperature-dependent avalanche 
correlation length $\ell(T)$ growing as $T$ is decreased.
Yet, there is a breakdown for this scaling, which could be 
naively related with the intrinsic avalanche length scale 
becoming comparable to the system size, $\ell(T)\sim L$, 
in the limit $T\to 0$, but that requires a more complex
argument at finite $T$.
Ref.~\cite{TaheiPRX2023} bases it on the scaling with $N$ of 
the difference between the lowest and second-lowest activation 
energies, leading to the aforementioned scaling breakdown size
$N_T\sim T^{-1/\delta}$.
}
\resub{
If we try to build a straightforward relationship and propose 
$T_c(N)\sim N^{-\delta}~L^{-d\delta}$, our resulting effective 
exponents do not coincide with those from~\cite{TaheiPRX2023}.
Therefore, we tend to believe that at a true finite temperature
dynamics the two-weakest activation energies argument might 
not be sufficient.
}

\resub{
Despite it being as a finite size effect,
$T_c(L)$ is something different from an intrinsic avalanche 
length scale becoming comparable to the system size, i.e., not simply $\ell(T_c)\sim L$, and is even not the same as the temperature that one could define from $N_T \sim T^{-1/\delta}$.
It seems to be more controlled by multi-nucleation and an increasing 
``frequency'' of plastic events, similar to what is discussed 
in~\cite{TakahaPRE2025}, and in the following we revise this 
line of thought again in our own words.
}
Essentially, a fixed $x_0$ or $T$ provides a fixed probability 
distribution ($x$ dependent) of activating sites at each time. 
In large systems this gives rise to multiple avalanche `seeds' 
taking place simultaneously in different part of the system.
This is, while one avalanche is evolving in a region, a new nucleus 
happens somewhere else and starts an avalanche growth. 
Since the definition of an avalanche is that it lasts while there is 
at least one active site in the system, these in principle uncorrelated 
nuclei end up being part of the `same' avalanche and therefore 
extending its life.
In the language of~\cite{TaheiPRX2023}, the multiple nucleus occurrence 
should act as imposing a ``cutoff by avalanches'' to the avalanche growth. 
That would be possible to consider if we had a way to distinguish causality
in different simultaneous growing nucleus, and to discriminate separated 
simultaneous avalanches happening in the system.
But the truth is that, in a real fixed temperature $T$ dynamics, those 
avalanches simply merge in a larger one and do not `cutoff' each other. 
Nevertheless, a length scale of such dynamics at finite 
temperature is provided by \resub{$\chi_4^{\tt peak}$ and, so far 
for the evidence, it seems to display} a maximum around $T=T_c$.
\resub{So, this is} a direction that deserves further exploration and 
conceptual discussion.

\subsection{Finite-size scaling at a fixed temperature $T<T_c$}

\begin{figure}[t!]
\includegraphics[width=0.9\columnwidth]{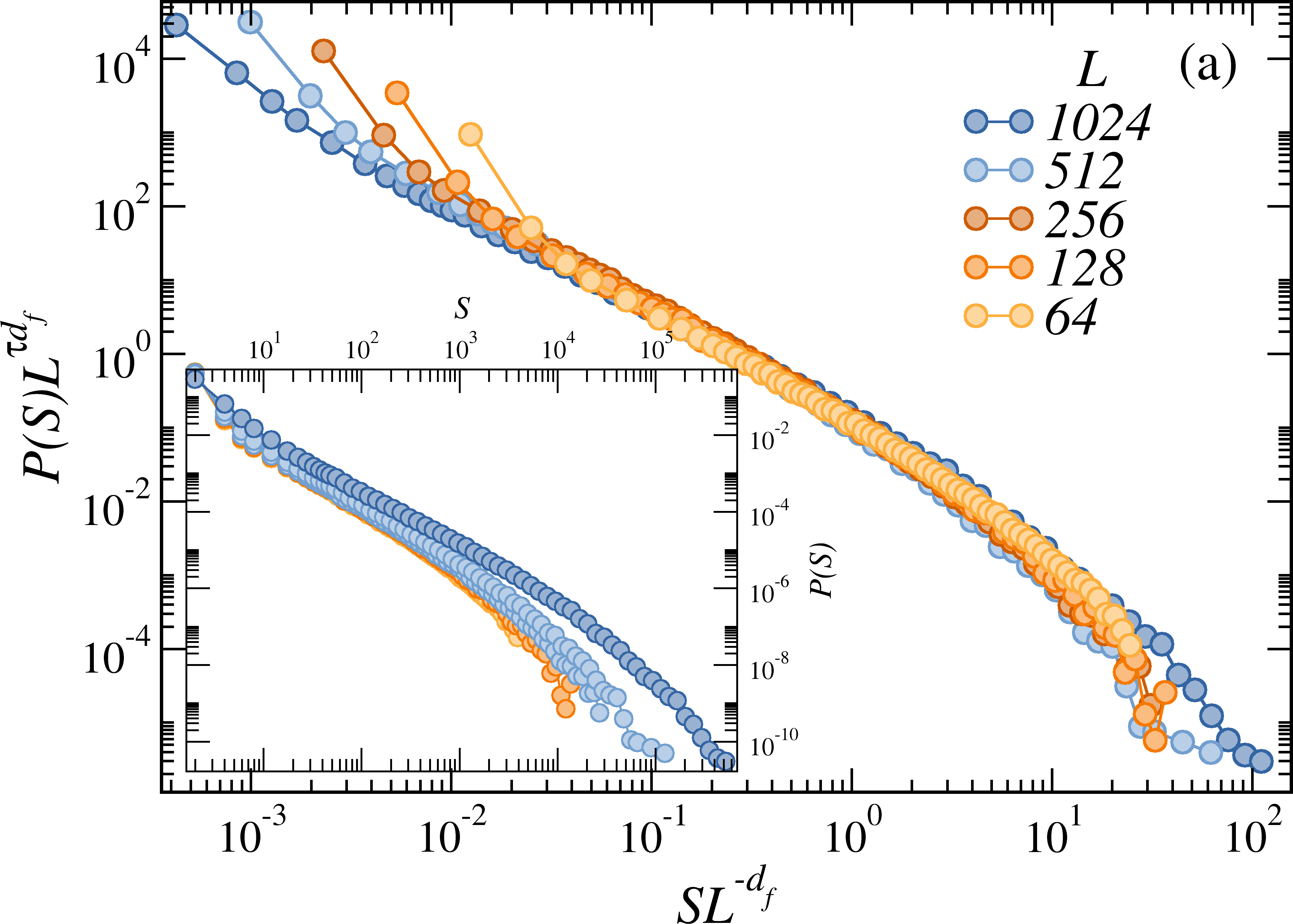}
\includegraphics[width=0.9\columnwidth]{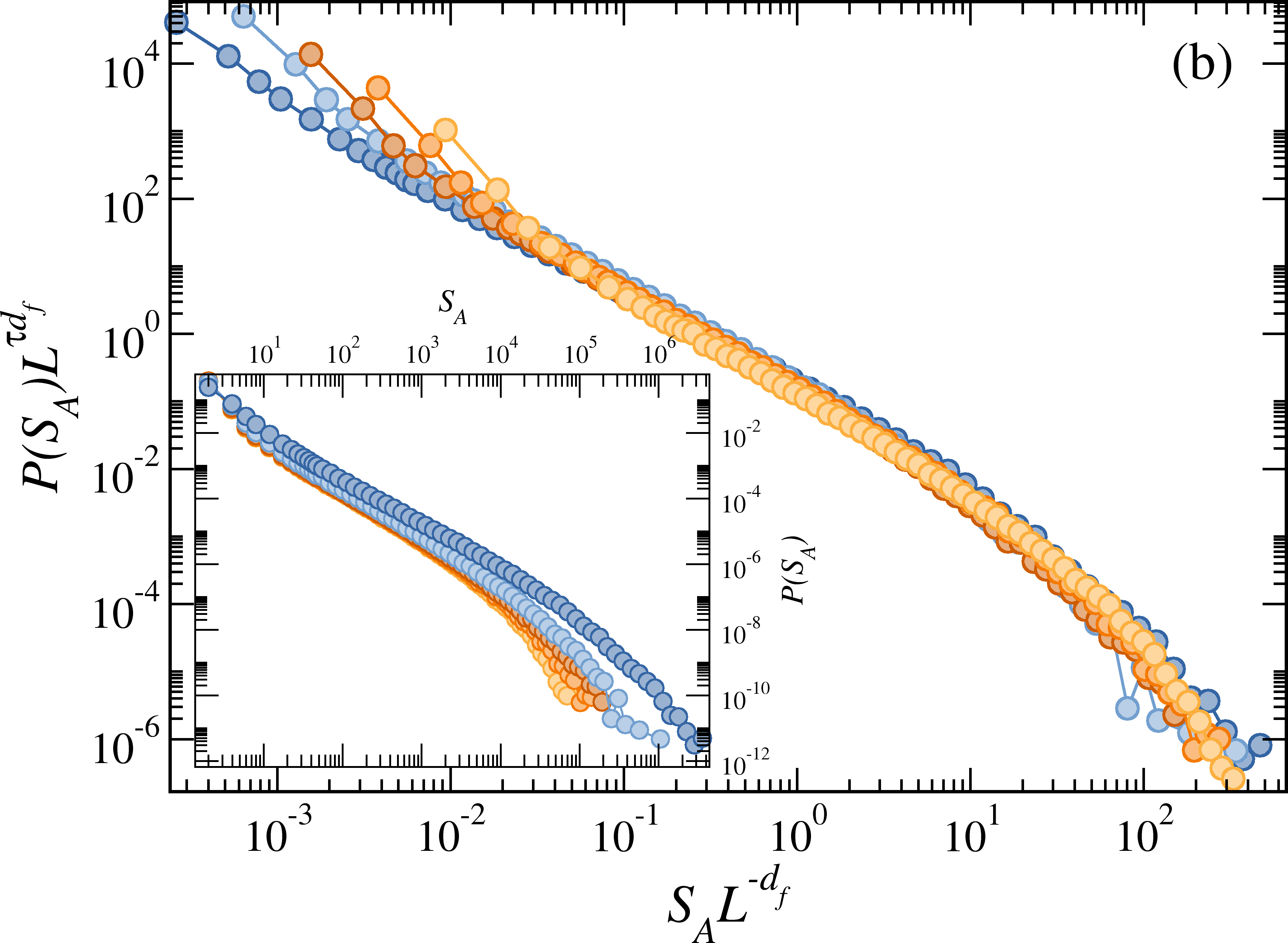}
\caption{\label{fig:p_s_sa_T0.0025_sizediff}
The main plots show the distributions of avalanche size and duration for various 
system sizes ($L=$ [64, 128, 256, 512, 1024]) at temperature $T=$ 0.0025:
(a) Probability distribution of the number of sites involved in an avalanche,
$P(S)L^{\tau d_{f}}$ vs $SL^{-d_{f}}$, with $d_{f}= 1.22$.
(b) Probability distribution of the total number of activations during 
an avalanche, $P(S_A)L^{\tau d_{f}}$ vs  $S_{A}L^{d_{f}}$, $d^{\prime}=1.29$.
The subpanels in each figure display the unscaled curves (raw data).
}
\end{figure}

We want now to study the effect of varying system size at fixed temperature.
We choose a small enough temperature, $T=0.0025$, such that $T<T_c(L)$ for all 
the range of studied system sizes (remember that $T_c(L)\to0$ as $L\to\infty$). 
Figure~\ref{fig:p_s_sa_T0.0025_sizediff} shows the results of a finite-size scaling for
$P(S)$ and $P(S_A)$.
Curves collapse can be obtained with the classical scaling form~\cite{LinPNAS2014,liu2016,JaglaPhysRevE2017}
\begin{equation}
    P(S)\propto S^{-\tau} g\left(\frac{S}{S_{c}}\right)
    \label{Eq.colapse_s}
\end{equation}
(and similarly for $S_A$).
The function $g\left(x\right)$ is a rapidly decaying function and $S_c$ denotes 
the cutoff scale. 
One proposes that this cutoff is dependent on the system size as $S_c \propto L^{d_f}$, 
where $d_f$ is a fractal dimension,
and then a master-curve can be found by plotting 
$P(S)L^{\tau d_f}$ vs. $S/L^{d_f}$.
Figures~\ref{fig:p_s_sa_T0.0025_sizediff}(a)-(b) show that good collapses
can be obtained for both $P(S)$ and $P(S_A)$ curves with $d_f\simeq 1.22$
and $d_f\simeq 1.29$ for $S$ and $S_A$, respectively.  
It should be noticed, nevertheless that these collapses are achieved in 
a nonstandard manner, since $\tau$ itself depends on $L$.
The values of $\tau$ that we estimate range from $\tau\simeq 1.5$ at $L=64$ 
to $\tau\simeq 1.35$ at $L=1024$, roughly alike for both $P(S)$ and $P(S_A)$.
More in detail, what we observe is that: up to $L=512$ $\tau$ suffers little
changes, but as size increases further, it shows more variation, decreasing 
from $\tau\sim 1.5$ to $\tau\sim 1.35$.
A way to justify this is by recalling that, having chosen a fixed temperature 
for the analysis, increasing $L$ brings us closer to the critical temperature 
(since $T_c(L)$ decreases with $L$, Fig.~\ref{fig:Tc_sizediff}), and the frequency 
of longer-lived avalanches increases, as we discussed around 
Fig.~\ref{fig:thermal_avalanches} in Sec.~\ref{sec:avalanchedistributions}.
In our largest system, even different scale-free regimes seem to emerge.
One can estimate in the distributions two power-law regions with 
different exponents (see Fig.~S6 for $L = 2048$ at $T=T_c = 0.0025 $
in Supp. Mat.~\cite{SM}).
A finite size scaling for avalanche duration distributions $P(D)$ 
at fixed $T$ can be found in Fig.~S9 in the Supp. Mat.~\cite{SM}.

\subsection{Avalanche size-duration mean relations}

\begin{figure}[t!]
\includegraphics[width=0.9\columnwidth]{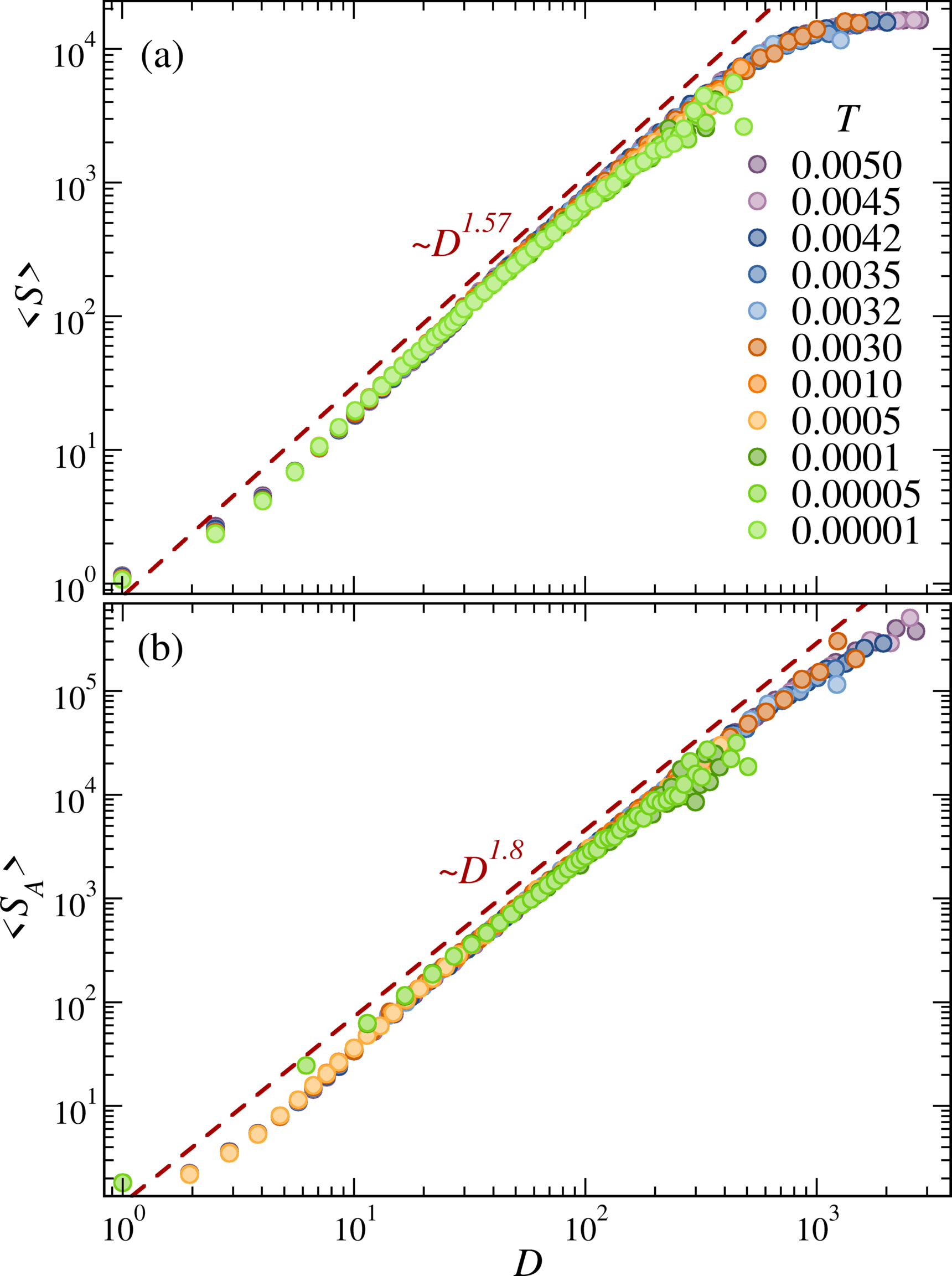}
\caption{\label{fig:mean_s_vs_T_TempsL128}
\textit{Avalanche size $\left< S \right>$ and $\left< S_{A} \right>$ vs. duration $D$}. 
Panel (a) shows $\left< S \right>$ and panel (b) shows $\left< S_{A} \right>$.
Different colored symbols correspond to different temperatures as indicated by the labels. 
Red dashed lines represent power-law fits. $\left<S\right> \propto D^{1.57}$
and $\left< S_{A} \right> \propto D^{1.8}$. 
Parameters: $L = 128$, $T \in [0.00001, 0.005].$ 
}
\end{figure}

We now examine the relation between mean avalanche size 
(be it $\left<S\right>$ or $\left<S_{A}\right>$) and duration ($D$).
Figure~\ref{fig:mean_s_vs_T_TempsL128} shows that, for different temperatures, 
curves lie on top of each other without the need of a scaling collapse.

For $\left<S\right>$, we first observe $\left<S\right> \propto D^{\delta}$ 
with $\delta \simeq 1.57$. 
At larger $D$,  $\left<S\right>$ saturates at the system size $N=128^2$ 
(only reached for the larger temperatures).
The estimated scaling $\left<S\right> \propto D^{1.57}$ turns out to be consistent 
with the size-duration relation found for driven athermal systems in~\cite{liu2016},
where the exponent $\delta$ was even shown to be almost independent on the dimension 
($d=2$ or $d=3$).
Even though this could simply be a coincidence, it's noteworthy;
let us make here a digression. 
In general, one defines a characteristic length scale $\ell$ to describe 
the scaling behavior.
One says that avalanches sizes scale as $S\sim \ell^{d_f}$, with $d_f$ 
a fractal dimension, and their duration as $D \sim \ell^z$, with $z$ 
the dynamical exponent.
Therefore, $S$ and $D$ are related by $S \sim \ell^z \sim D^{d_f/z}$.
In other words, $\delta=d_f/z$.
The similarity of $\delta$ between sheared and thermally activated systems 
suggests that the ratio $d_f/z$ is robust to the triggering mechanism.
Moreover, we can extract $z$ from here. 
We know that $d_f\simeq 1$ in the case of 2D sheared systems, 
where an imposed direction fixes the angular orientation of the propagator.
In our current thermally activated system, with randomized orientation 
propagator, $d_f\simeq 1.22$ in two-dimensions.
So, $z$ should be also larger (approx. $z\simeq 1.92$)
than the one measured in~\cite{liu2016} in order to yield 
a similar ratio.

For $\left<S_A\right>$ we observe a law $\left<S_A\right> \propto D^{\delta}$ with 
$\delta \simeq 1.8$, with only a weak bending, which does not even develop 
in a saturation, at the largest durations.
$\left<S_{A}\right>$ grows with $D$ faster than $\left<S\right>$. 
This means that, even for relatively brief and small avalanches 
(much before system-size saturation), there are multiple activations 
of at least some of the sites involved in them.
In the power-law regime, the relation is $\left<S_A\right> \sim \left<S\right>^{1.15}$.
Interestingly, the estimated value of $\delta \simeq 1.8$ for $\left<S_A\right>$
aligns well with the findings of Budrikis \textit{et al.} in 2D models of amorphous 
materials~\cite{budrikis2013,budrikis2017}.

A finite size analysis of $\left<S_{A}\right>$ grows with $D$
performed at $T=T_c$ can be found in Fig.~S7 in the Supp. Mat.~\cite{SM}.

\subsection{Probability distributions of $x_{\text{min}}$ at finite temperature}

\begin{figure}[t!]
\includegraphics[width=0.47\textwidth]{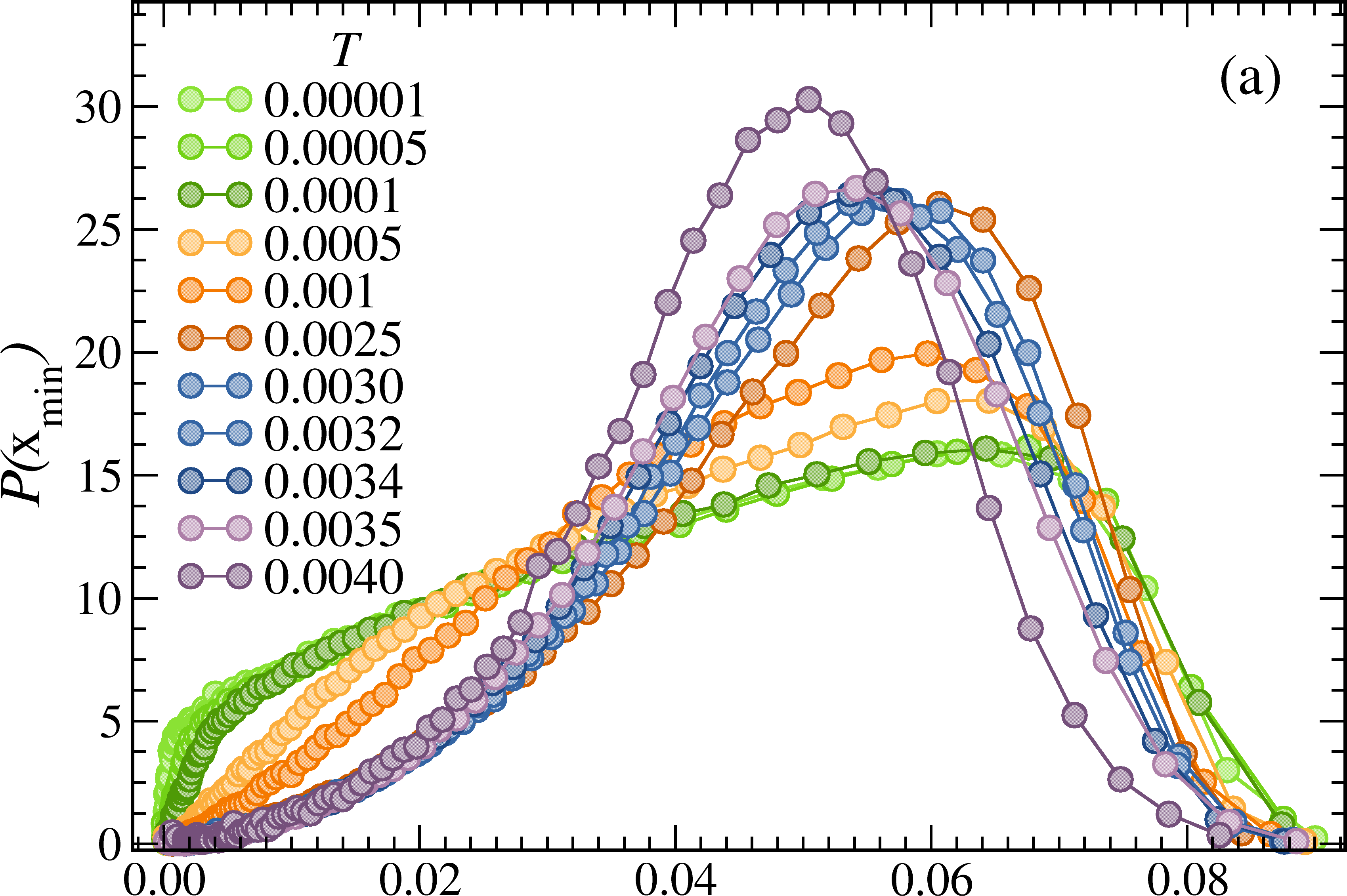}
\includegraphics[width=0.47\textwidth]{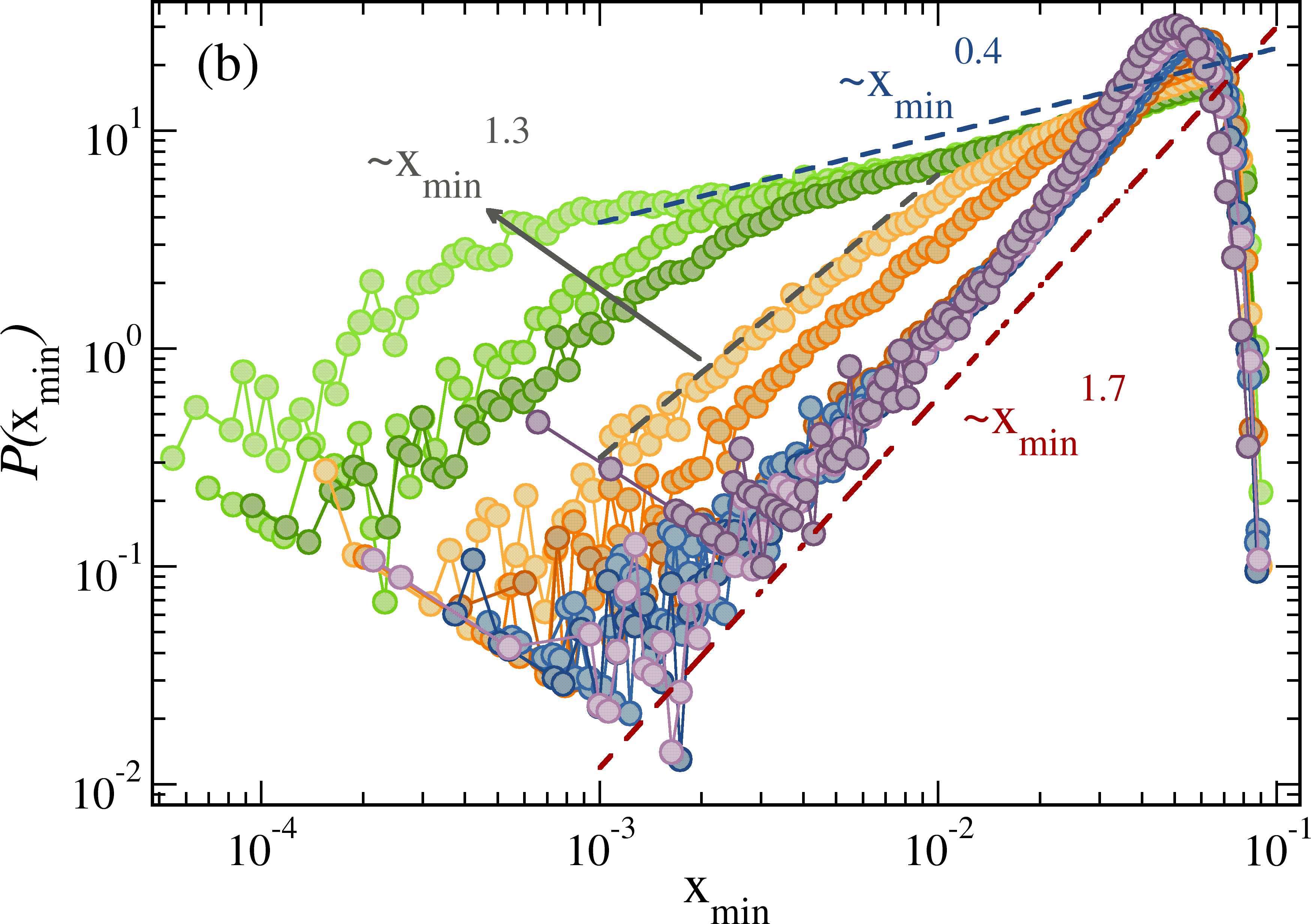}
\caption{ 
The distribution curves of $x_{\text{min}}$ [$P(x_{\text{min}})$] for a system size of $L = 512$ are depicted at various temperatures ($T =$ [0.00001, 0.00005, 0.0001, 0.0005, 0.0010, 0.0025, 0.0030, 0.0032, 0.0034, 0.0035, 0.0040]). Panel (a) illustrates $P(x_{\text{min}})$ on a linear–linear scale, whereas panel (b) presents the same data using a log–log scale. The red dashed lines denote power-law behaviors corresponding to distinct temperature regimes. As the temperature increases, three distinct power-law regimes are observed: $x_{\text{min}}^{0.4}$ for low temperatures, $x_{\text{min}}^{1.3}$ for intermediate temperatures, and $x_{\text{min}}^{1.7}$ for high temperatures.
}\label{fig:Xmin_FixSize}
\end{figure}

In sheared amorphous systems, avalanche statistics are tightly connected to 
the distribution of local distances to yielding, $P(x)$, which acts as a 
fingerprint of the yielding transition~\cite{LinEPL2014,LinPNAS2014,liu2016,KarimiPhysRevE2017,FerreroSM2019,FerreroJPCM2021}.
Here we analyze the analogous quantity for thermally activated avalanches.

We define the local distance to yielding at site $i$ as
$x_i = \sigmaY_i - \sigma_i$, where $\sigma_i$ is the local stress and $\sigmaY_i$ the local threshold.
In our model each site has two symmetric thresholds $\pm\sigmaY$, identical for all $i$.
Accordingly, a unique distance can be written as
\begin{equation}
x_i = \min\{(\sigmaY - \sigma_i),(\sigma_i + \sigmaY)\}.
\end{equation}

We focus on the minimum distance
$x_{\text{min}}=\min_i x_i$,
evaluated immediately after an avalanche ends.
Sampling over many events allows us to construct the distribution $P(x_{\text{min}})$.

Figure~\ref{fig:Xmin_FixSize} shows $P(x_{\text{min}})$ for a fix system 
size and different temperatures; panel (a) shows the distributions in 
lin-lin scale and panel (b) shows the same data in log-log.
The first thing to notice is that $P(x_{\text{min}})$ changes considerably
with varying temperature, exhibiting distinct behaviors at different 
$x_{\text{min}}$ ranges.
First, in Fig.~\ref{fig:Xmin_FixSize}(a) we observe how the `gap' in 
$P(x_{\text{min}})$ is opened as temperature increases, in coincidence 
with what is discussed when increasing the threshold $x_0$ in the extremal 
dynamics protocol of~\cite{TaheiPRX2023}.
Notice that when reaching $T_c(L)\simeq0.0034$ the gap saturates.
Even more, the whole $P(x_{\tt min})$ distribution looks very little
perturbed and curves pretty much overlap around $T_c(L)$.
If we keep increasing the temperature beyond $T_c(L)$ eventually the 
distribution $P(x_{\text{min}})$ changes the form of its peak, which
also shifts to smaller values, but the lower tail of $P(x_{\text{min}})$ 
remains approximately unchanged.
Second, let us analyze the functional behavior of $P(x_{\text{min}})$
at small $x_{\text{min}}$.
Following the same principles that lead to the discussion of $P(x)$ in driven amorphous systems~\cite{LinEPL2014,LinPNAS2014,liu2016,KarimiPhysRevE2017,FerreroSM2019,FerreroJPCM2021},
one expects a similar boundary effect at the local threshold $x=0$, namely a power-law departure of the distribution from that boundary~\cite{TaheiPRX2023}.
If this holds for $P(x)$, it should also hold for $P(x_{\text{min}})$ (although the converse is not necessarily true~\cite{FerreroJPCM2021}).
We therefore propose
\begin{equation}
P(x_{\text{min}}) \sim x_{\text{min}}^{\theta}
\end{equation}
at small $x_{\text{min}}$.

Around $T=T_c$ we measure a rather large exponent, $\theta \simeq 1.7$, as shown in Fig.~\ref{fig:Xmin_FixSize}(b).
While the absence of external driving allows $\theta$ to deviate from the small values typical of driven yielding~\cite{FerreroJPCM2021, liu2016}, a value $\theta>1$ cannot be explained by independent local rearrangements.
For instance, a simple random-walk process with an absorbing boundary at $x=0$ yields $P(x)\sim x$, i.e. $\theta=1$.
More generally, for a fractional Brownian motion (fBm) with Hurst exponent $H$, one expects~\cite{FerreroJPCM2021,LinPRX2016}
\begin{equation}
\theta=\frac{1}{2H}.
\end{equation}
Thus, $\theta>1$ implies $H<0.5$, corresponding to a process with negatively correlated increments.
To account for $\theta\simeq1.7$ one would need $H\simeq0.3$.
It's like a fBm with a tendency to quickly revert to it's mean previous state.

Interestingly, similarly large values of $\theta$ have been reported (even by one of us) 
in the context of inertial avalanches~\cite{KarimiPhysRevE2017}, where long-lasting 
events progressively exhaust small energy barriers, effectively opening a gap in 
$P(x)$ near $x=0$ and steepening the distribution.
A finite temperature may play a comparable role in the exhaustion of small energetic 
barriers if avalanches persist long enough to act and eliminate them.
However, this analogy may reflect a consequence rather than the fundamental origin of $\theta>1$, since overdamped driven systems systematically display $\theta<1$.

As temperature is decreased, we observe in Fig.~\ref{fig:Xmin_FixSize}(b)
that another regime on the distribution emerges.
For the lowest temperatures the behavior is characterized by the power-law $\sim x_{\text{min}}^{0.4}$ at intermediate values of $ x_{\text{min}}$.
This is consistent with a driven-yielding-like behavior for the avalanche
distributions, as we observe for low temperatures in Fig.~\ref{fig:thermal_avalanches}.
Interestingly, though, a fastest decrease with a $\theta>1$ still persist 
close enough $x=0$.
This can be seen as a feature of thermal systems, since in driving yielding the
situation is always the opposite one: the occurrence of a `plateau'~\cite{TyukodiPRE2019, RuscherSM2020, FerreroJPCM2021, KorchinskiPRE2021} when $x\to 0$.

We have also computed $\langle x_{\min} \rangle$ as a function of 
temperature and system size.
As $L$ increases, we observe a power-law decrease
$\langle x_{\min} \rangle \sim L^{-0.16}$ (see Fig.~S10 in the 
Supp. Mat.~\cite{SM}), with no clear sign of saturation up to $L=2048$.
For thermally activated \emph{driven} systems in the intermittent regime,
Korchinski \textit{et al.}~\cite{KorchinskiSM2024}
predicted
\begin{equation}
\langle x_{\min} \rangle \sim \left( x_{\mathrm{a}}^{\,1+\theta} + A L^{-d} \right)^{\frac{1}{1+\theta}},
\end{equation}
which implies saturation to the thermal activation scale 
$x_{\mathrm{a}}$ as $L\to\infty$ at fixed $T$.
The absence of such saturation in our data suggests that the mechanism 
controlling $x_{\min}$ in our avalanche-triggered protocol differs 
from the activation-controlled plateau reported for steadily driven systems.

\section{Discussion}

We have analyzed the emergence of purely thermal dynamical heterogeneities 
and avalanche activity in elastoplastic models of amorphous solids, 
identifying a finite-size-dependent temperature scale that organizes the 
dynamics.

Dynamical heterogeneities, quantified through the four-point susceptibility 
$\chi_4$ extracted from persistence measurements, reveal a twofold behavior. 
As $T\to0$, the peak of $\chi_4$ grows, consistently with a growing dynamical 
correlation length in the glassy regime. 
At the same time, $\chi_4$ exhibits a maximum at a finite temperature $T^*$, 
which we have identified with a crossover temperature $T_c(L)$ separating 
distinct dynamical regimes.

Avalanche statistics further clarify this separation. 
For $T<T_c(L)$, avalanche sizes and durations display power-law distributions 
with exponential cutoffs, characteristic of intermittent dynamics. 
For $T>T_c(L)$, the distributions develop pronounced shoulders or peaks, 
signaling the emergence of runaway avalanches and the onset of 
thermally assisted flow. 
Close to and above $T_c(L)$, avalanche duration provides a natural criterion 
to distinguish between scale-free intermittent events and system-spanning 
runaway avalanches that remain active for stochastic lifetimes.

A central result is that $T_c$ depends on system size and decreases as 
$L$ increases, with $T_c \propto L^{-0.23} \propto N^{-0.115}$. 
Our data suggest that $T_c(L)\to 0$ as $L\to\infty$, albeit with a slow 
decay. 
Within our model and approximations, this behavior is compatible 
with the scenario in which arbitrarily small but finite temperatures 
ultimately destabilize the intermittent regime in sufficiently large systems. 
Whether this extrapolation survives beyond the present framework remains 
an open question.

Our results also clarify the interpretation of extremal dynamics protocols. 
Previously reported extremal-dynamics results are not generically equivalent 
to a strict $T=0^+$ limit, but rather correspond to an effective finite 
temperature controlled by the imposed threshold $x_0=x_c$. 
The role of this threshold is analogous to that of $T_c(L)$ in the 
fully thermal protocol: above it, the system develops runaway avalanches 
and crosses over to a fluidized regime. 
An important difference, however, is that in the present fully thermal 
dynamics, the subcritical regime $T<T_c(L)$ exhibits temperature-dependent 
power-law statistics, revealing a richer structure than that captured by 
extremal dynamics alone.
These results further show that thermal activation alone can 
reorganize marginal stability and avalanche dynamics in amorphous solids,
introducing a finite-size-controlled instability scale that is absent 
in strictly driven overdamped systems.

More broadly, this mechanism may extend to other disordered elastic systems, 
including interfaces in depinning problems, where thermal fluctuations 
compete with collective elastic interactions. 
Although equilibrium properties and thermally assisted creep under small 
driving are well understood in that context, the possibility that purely 
thermal fluctuations alone could reorganize marginal stability and generate
avalanche-like activity in the absence of external forcing has received 
comparatively little attention. 
Exploring this limit could uncover analogous instability scales and 
broaden the understanding of thermally activated dynamics in marginally 
stable systems.

\begin{acknowledgments} 
We would like to thank fruitful discussions with E.A.~Jagla 
and A.~Rosso and useful feedback from M. Ozawa.
G. R.-L. would like to thank A. B. Kolton, from Instituto Balseiro,
for helpful discussions on the subject and for technical support with the CUDA implementation.
G.R.-L. wants to thank Prof. Nicolás García for his support.
We acknowledge financial support from PIP 2021-2023 CONICET Project Nº~757
and the CNRS IRP Project ``Statistical Physics of Materials''.
G.R.-L. acknowledges support from Universidad Nacional del Sur 
and FONCyT grant PICT-2021-I-A-01272. 
We acknowledge computational resources provided by the 
HPC cluster of the Physics Department at Centro Atómico Bariloche
and the computing cluster at IFISUR, Bahía Blanca.
\end{acknowledgments}

\bibliographystyle{apsrev4-2}

%

\end{document}